\documentclass[12pt]{article}

\usepackage{amsmath}
\usepackage{amssymb}
\usepackage{amsfonts}
\usepackage{latexsym}
\usepackage{color}

\catcode `\@=11 \@addtoreset{equation}{section}

\catcode `\@=12

\newcommand{\be}{\begin{equation}}
\newcommand{\en}{\end{equation}}
\newcommand{\bea}{\begin{eqnarray}}
\newcommand{\ena}{\end{eqnarray}}
\newcommand{\beano}{\begin{eqnarray*}}
\newcommand{\enano}{\end{eqnarray*}}
\newcommand{\bee}{\begin{enumerate}}
\newcommand{\ene}{\end{enumerate}}

\newcommand{\N}{\mathfrak N}

\newcommand{\mc}{\mathcal}

\newcommand{\D}{{\mc D}}

\newcommand{\E}{{\cal E}}
\newcommand{\F}{{\cal F}}

\newcommand{\Lc}{{\cal L}}

\newcommand{\1}{1 \!\! 1}

\newcommand{\PP}{\mc P}

\newcommand{\Hil}{\mc H}

\newtheorem{thm}{Theorem}

\catcode `\@=11 \@addtoreset{equation}{section}
\catcode `\@=12

\begin{document}

\thispagestyle{empty}

\vspace*{2cm}

\begin{center}
{\Large \bf  Pseudo-bosons, so far}\\[10mm]

{\large F. Bagarello}\\
  Dipartimento di Metodi e Modelli Matematici,
Facolt\`a di Ingegneria,\\ Universit\`a di Palermo, I-90128  Palermo, Italy\\
e-mail: bagarell@unipa.it\\ Home page:
www.unipa.it/$^\sim$bagarell\\

\vspace{3mm}

\end{center}

\vspace*{2cm}

\begin{abstract}
\noindent  In the past years several extensions of the canonical commutation relations have been proposed by different people in different contexts and some interesting physics and mathematics have been deduced. Here, we review some recent results on the so-called {\em
pseudo-bosons}. They arise from a special deformation of the
canonical commutation relation $[a,a^\dagger]=\1$, which is
replaced by $[a,b]=\1$, with $b$ not necessarily equal to $a^\dagger$. We start discussing some of their mathematical properties and then we discuss several examples.

\end{abstract}

\vspace{2cm}



\vfill


\newpage

\section{Introduction}

In the past years several extensions  of the canonical (anti-)commutation relations have been proposed by different people in different contexts and some interesting physics and mathematics have been deduced. These have produced several kind of {\em quasi-particles}, among which we can cite anyons (more on a physical side), see \cite{wil} and references therein, and quons (more on a mathematical side), \cite{moh}.

It is not our aim to review here all these extensions and their related (quasi-)particles. In fact, we are here interested in considering a very particular modification of the canonical commutation rule $[a,a^\dagger]=\1$, the one which, in our knowledge, Trifonov first in \cite{tri} called {\em pseudo-bosonic} commutation rule, and which looks like $[a,b]=\1$, where $b$ is not required to be equal to $a^\dagger$.

It should be mentioned that the words {\em pseudo-bosons} have been used years before Trifonov, but with a different meaning or in different contexts with respect to our interests, \cite{other1}. On the opposite side, pseudo-bosons were already found several years before Trifonov by many authors, \cite{other2}, but they were not called this way and the analysis which was later proposed by us was not considered.

In a series of recent papers \cite{bagpb1,bagpb2,bagpb3,bagcal}, we have investigated some mathematical aspects of these
 pseudo-bosons.  We have
shown that, under suitable assumptions, $N=ba$ and $N^\dagger=a^\dagger b^\dagger$ can be both
diagonalized, and that their spectra coincide with the set of
natural numbers (including 0), ${\Bbb N}_0$. However the two sets of
related eigenvectors are not orthonormal (o.n) bases but,
nevertheless, they are automatically { biorthogonal\/}. In most of the
examples considered so far, they are bases of the Hilbert space of the system,
$\Hil$, and, in some cases, they turn out to be {\em Riesz bases\/}.

In \cite{bagpb4} and \cite{abg} some physical examples arising from  quantum mechanics have been discussed. In particular, a difference between what we have called {\em regular pseudo-bosons} and {\em pseudo-bosons} has been introduced, to better focus on the mathematical or on the physical aspects of these {\em particles}.

As already stressed, this paper is not meant to review all the results related in different way to pseudo-bosons of all kind, but only those obtained so far by the present author and his collaborators, and is organized as follows:  in the next section we
introduce and discuss $d$-dimensional pseudo-bosons analyzing some
of their mathematical properties and their related coherent
states.  In Sections III-IX we discuss several examples in the context of Section II.
Section X contains our conclusions.

\section{The commutation rules}

In this section we will discuss a $d$-dimensional version  of
what originally proposed in \cite{bagpb1}.

Let $\Hil$ be a given Hilbert space with scalar product
$\left<.,.\right>$ and related norm $\|.\|$. We introduce $d$
pairs of operators, $a_j$ and $b_j$, $j=1,2,\ldots,d$, acting on $\Hil$ and
satisfying the following commutation rules \be [a_j,b_j]=\1,
\label{21} \en where $j=1,2,\ldots,d$, while all the other commutators are zero. Of course, these collapse to the CCR's for $d$
independent modes if $b_j=a^\dagger_j$, $j=1,2,\ldots,d$. It is well known
that $a_j$ and $b_j$ are unbounded operators, so they cannot be
defined on all of $\Hil$. Following \cite{bagpb1}, and writing
$D^\infty(X):=\cap_{p\geq0}D(X^p)$ (the common  domain of all the powers of the
operator $X$), we consider the
following:

\vspace{2mm}

{\bf Assumption 1.--} there exists a non-zero
$\varphi_{\bf 0}\in\Hil$ such that $a_j\varphi_{\bf 0}=0$, $j=1,2,\ldots,d$,
and $\varphi_{\bf 0}\in D^\infty(b_1)\cap D^\infty(b_2)\cap\cdots\cap D^\infty(b_d)$.

{\bf Assumption 2.--} there exists a non-zero $\Psi_{\bf 0}\in\Hil$
such that $b_j^\dagger\Psi_{\bf 0}=0$, $j=1,2,\ldots,d$, and $\Psi_{\bf 0}\in
D^\infty(a_1^\dagger)\cap D^\infty(a_2^\dagger)\cap\cdots\cap D^\infty(a_d^\dagger)$.

\vspace{2mm}

Under these assumptions we can introduce the following vectors in
$\Hil$:

\be
\left\{
\begin{array}{ll}
\varphi_{\bf n}:=\varphi_{n_1,n_2,\ldots,n_d}=\frac{1}{\sqrt{n_1!n_2!\cdots n_d!}}\,b_1^{n_1}\,b_2^{n_2}\cdots b_d^{n_d}\,\varphi_{\bf 0}\\
\Psi_{\bf n}:=\Psi_{n_1,n_2,\ldots,n_d}=\frac{1}{\sqrt{n_1!n_2!\cdots n_d!}}\,{a_1^\dagger}^{n_1}\,{a_2^\dagger}^{n_2}\cdots {a_d^\dagger}^{n_d}\,\Psi_{\bf 0},
\end{array}
\right.\label{22}\en $n_j=0, 1, 2,\ldots$ for all $j=1,2,\ldots,d$. Let us now define the unbounded
operators $N_j:=b_ja_j$ and $\N_j:=N_j^\dagger=a_j^\dagger
b_j^\dagger$, $j=1,2,\ldots,d$.  It is possible to check that
$\varphi_{\bf n}$ belongs to the domain of $N_j$, $D(N_j)$, and
$\Psi_{\bf n}\in D(\N_j)$, for all possible $\bf n$. Moreover,
\be N_j\varphi_{\bf n}=n_j\varphi_{\bf n},  \quad \N_j\Psi_{\bf n}=n_j\Psi_{\bf n}.
\label{23}\en

Under the above assumptions, and if we chose the normalization of
$\Psi_{\bf 0}$ and $\varphi_{\bf 0}$ in such a way that
$\left<\Psi_{\bf 0},\varphi_{\bf 0}\right>=1$, we find that \be
\left<\Psi_{\bf n},\varphi_{\bf m}\right>=\delta_{\bf n,m}=\prod_{j=1}^d \delta_{n_j,m_j}. \label{27}\en This means that the sets
$\F_\Psi=\{\Psi_{\bf n}\}$ and
$\F_\varphi=\{\varphi_{\bf n}\}$ are {\em biorthogonal} and,
because of this, the vectors of each set are linearly independent.
If we now call $\D_\varphi$ and $\D_\Psi$ respectively the linear
span of  $\F_\varphi$ and $\F_\Psi$, and $\Hil_\varphi$ and
$\Hil_\Psi$ their closures, then \be f=\sum_{\bf n}
\left<\Psi_{\bf n},f\right>\,\varphi_{\bf n}, \quad \forall
f\in\Hil_\varphi,\qquad  h=\sum_{\bf n}
\left<\varphi_{\bf n},h\right>\,\Psi_{\bf n}, \quad \forall
h\in\Hil_\Psi. \label{210}\en What is not in general ensured is
that
$\Hil_\varphi=\Hil_\Psi=\Hil$. Indeed, we can only state that
$\Hil_\varphi\subseteq\Hil$ and $\Hil_\Psi\subseteq\Hil$. However,
motivated by the examples discussed so far in the literature,  we consider

\vspace{2mm}

{\bf Assumption 3.--} The above Hilbert spaces all coincide:
$\Hil_\varphi=\Hil_\Psi=\Hil$.

\vspace{2mm}

This means, in particular,
that both $\F_\varphi$ and $\F_\Psi$ are bases of $\Hil$. Let us
now introduce the operators $S_\varphi$ and $S_\Psi$ via their
action respectively on  $\F_\Psi$ and $\F_\varphi$: \be
S_\varphi\Psi_{\bf n}=\varphi_{\bf n},\qquad
S_\Psi\varphi_{\bf n}=\Psi_{\bf n}, \label{213}\en for all $\bf n$, which also imply that
$\Psi_{\bf n}=(S_\Psi\,S_\varphi)\Psi_{\bf n}$ and
$\varphi_{\bf n}=(S_\varphi \,S_\Psi)\varphi_{\bf n}$, for all
$\bf n$. Hence \be S_\Psi\,S_\varphi=S_\varphi\,S_\Psi=\1 \quad
\Rightarrow \quad S_\Psi=S_\varphi^{-1}. \label{214}\en In other
words, both $S_\Psi$ and $S_\varphi$ are invertible and one is the
inverse of the other. Furthermore, we can also check that they are
both positive, well defined and symmetric, \cite{bagpb1}. Moreover, it is possible to write these operators in the
bra-ket notation as \be S_\varphi=\sum_{\bf n}\,
|\varphi_{\bf n}><\varphi_{\bf n}|,\qquad S_\Psi=\sum_{\bf n}
\,|\Psi_{\bf n}><\Psi_{\bf n}|. \label{212}\en
 These expressions are
only formal, at this stage, since the series may not converge in
the uniform topology and the operators $S_\varphi$ and $S_\Psi$ could be unbounded.
Indeed we know,  \cite{you}, that two biorthogonal bases are related by a bounded operator, with bounded inverse, if and only if they are Riesz bases\footnote{Recall that a set of vectors $\phi_1, \phi_2 , \phi_3 , \; \ldots \; ,$ is a Riesz basis of a Hilbert space $\mathcal H$, if there exists a bounded operator $V$, with bounded inverse, on $\mathcal H$, and an o.n. basis of $\Hil$,  $\varphi_1, \varphi_2 , \varphi_3 , \; \ldots \; ,$ such that $\phi_j=V\varphi_j$, for all $j=1, 2, 3,\ldots$}. This was our motivation  in \cite{bagpb1} to consider the following

\vspace{2mm}

{\bf Assumption 4.--} $\F_\varphi$ and $\F_\Psi$ are  both Riesz bases.

\vspace{2mm}

Therefore, as already stated, $S_\varphi$ and $S_\Psi$ are bounded operators and their domains  can be taken to be all of $\Hil$. While Assumptions 1, 2 and 3 are quite often satisfied, as the examples contained in the following sections prove,  it is quite difficult to find {\bf physical} examples satisfying also Assumption 4. On the other hand, it is rather easy to find {\bf mathematical} examples satisfying all the assumptions, see Section II.1. Hence we introduce a difference in the notation: we call {\em pseudo-bosons} (PB) those satisfying  the first three assumptions, while, if they also satisfy Assumption 4, they will be called {\em regular pseudo-bosons} (RPB).

As already discussed in our previous papers,  these
$d$-dimensional pseudo-bosons give rise to interesting
intertwining relations among non self-adjoint operators, see in particular
\cite{bagpb3} and references therein. For instance, it is easy to
check that \be S_\Psi\,N_j=\N_jS_\Psi \quad \mbox{ and }\quad
N_j\,S_\varphi=S_\varphi\,\N_j, \label{219}\en $j=1,2,\ldots,d$. This is
related to the fact that the eigenvalues of, say, $N_1$ and $\N_1$,
coincide and that their eigenvectors are related by the operators
$S_\varphi$ and $S_\Psi$, in agreement with the literature on
intertwining operators, \cite{intop,bag1}, and on pseudo-Hermitian
quantum mechanics, see \cite{mosta} and references therein.

\subsection{Construction of regular pseudo-bosons}

We will show here that each Riesz basis produces some regular pseudo-bosons. Let $\F_\varphi:=\{\varphi_{\bf n}\}$ be a Riesz basis of $\Hil$ with bounds $A$ and $B$, $0<A\leq B<\infty$: $$A\|f\|^2\leq\sum_{\bf n}\left|\left<f,\varphi_{\bf n}\right>\right|^2\leq B\|f\|^2, $$ for all $f\in\Hil$. The associated frame operator $S:=\sum_{\bf n}\,|\varphi_{\bf n}><\varphi_{\bf n}|$ is bounded, positive and admits a bounded inverse, \cite{chri}. The set $\F_{\hat\varphi}:=\{\hat\varphi_{\bf n}:=S^{-1/2}\varphi_{\bf n}\}$ is an o.n. basis of $\Hil$. Hence we can define $d$ lowering operators $a_{j,\hat\varphi}$ on  $\F_{\hat\varphi}$ as $a_{j,\hat\varphi}\hat\varphi_{\bf n}=\sqrt{n_j}\,\hat\varphi_{\bf n_{j-}}$, and their adjoints, $a_{j,\hat\varphi}^\dagger$, as $a_{j,\hat\varphi}^\dagger\hat\varphi_{\bf n}=\sqrt{n_j+1}\,\hat\varphi_{\bf n_{j+}}$. Here ${\bf n}_{j-}=(n_1,\ldots,n_j-1,\ldots,n_d)$ and ${\bf n}_{j+}=(n_1,\ldots,n_j+1,\ldots,n_d)$. Hence $[a_{j,\hat\varphi},a_{k,\hat\varphi}^\dagger]=\delta_{j,k}\,\1$. Notice that our notation here makes explicit the fact that the raising and lowering operators depend on $\F_{\hat\varphi}$ and, therefore, on $\F_\varphi$. If we now define $a_j:=S^{1/2}\,a_{j,\hat\varphi}\,S^{-1/2}$, this acts on the original set $\F_\varphi$ as a lowering operator. However, since $\F_\varphi$ is not an o.n. basis in general, $a_j^\dagger$ is not a raising operator for $\F_\varphi$, and $[a_j,a_k^\dagger]\neq\delta_{j,k}\,\1$. If we now define the operator $b_j:=S^{1/2}\,a_{j,\hat\varphi}^\dagger\,S^{-1/2}$, it is clear that in general  $b_j\neq a_j^\dagger$. Moreover, $b_j$ acts on $\varphi_{\bf n}$ as a raising operator: $b_j\,\varphi_{\bf n}=\sqrt{n_j+1}\,\varphi_{{\bf n}_{j+}}$, for all $\bf n$, and we also have $[a_j,b_k]=\delta_{j,k}\,\1$. So we have constructed two operators satisfying (\ref{21}) and which are not related by a simple conjugation. This is not the end of the story. Indeed we can check that:
\begin{enumerate}
\item Assumption 1 is verified since $\varphi_{\bf 0}$ is annihilated by $a_j$ and belongs to the domain of all the powers of $b_j$.
\item As for Assumption 2, it is enough to define $\Psi_{\bf 0}=S^{-1}\,\varphi_{\bf 0}$. With this definition $b_j^\dagger\,\Psi_{\bf 0}=0$ and $\Psi_{\bf 0}$ belongs to the domain of all the powers of $a_j^\dagger$.
\item Since $\F_\varphi$ is a Riesz basis of $\Hil$ by assumption, then $\Hil_\varphi=\Hil$. Moreover, \cite{bagpb1}, the vectors $\Psi_{\bf n}$  can be written as $\Psi_{\bf n}=S^{-1}\,\varphi_{\bf n}$, for all $\bf n$. Hence $\F_\Psi$ is in duality with $\F_\varphi$ and therefore is a Riesz basis of $\Hil$ as well, \cite{chri}. Hence $\Hil_\Psi=\Hil$. This proves Assumption 3.
\item As for Assumption 4, this is exactly the hypothesis originally assumed here, i.e. that $\F_\varphi$ is a Riesz basis.
\end{enumerate}

\subsection{Coherent states}

As it is well known there exist several different, and not always
equivalent, ways to define {\em
coherent states}, \cite{book1,book2}. In this paper we will
adopt the following definition,  generalizing what we did in
\cite{bagpb1}. Let $z_j$, $j=1,2,\ldots,d$ be $d$ complex variables, $z_j\in \D$ (some common domain in $\Bbb{C}$), and let us introduce the following
operators: \be
U_j(z_j)=e^{z_j\,b_j-\overline{z}_j\,a_j}=e^{-|z_j|^2/2}\,e^{z_j\,b_j}\,e^{-\overline{z}_j\,a_j},
\quad
V_j(z_j)=e^{z_j\,a_j^\dagger-\overline{z}_j\,b_j^\dagger}=e^{-|z_j|^2/2}\,e^{z_j\,a_j^\dagger}\,e^{-\overline{z}_j\,b_j^\dagger},
\label{31x}\en $j=1,2,\ldots,d$,  \be
U(z_1,z_2,\ldots,z_d):=U_1(z_1)\,U_2(z_2)\,\cdots\,U_d(z_d),\qquad
V(z_1,z_2,\ldots,z_d):=V_1(z_1)\,V_2(z_2)\,\cdots\,V_d(z_d), \label{31b}\en and the following
vectors: \be \varphi(z_1,z_2,\ldots,z_d)=U(z_1,z_2,\ldots,z_d)\varphi_{\bf 0},\qquad
\Psi(z_1,z_2,\ldots,z_d)=V(z_1,z_2,\ldots,z_d)\,\Psi_{\bf 0}. \label{32}\en \vspace{2mm}

{\bf Remarks:--} (1) Due to the commutation rules for the
operators $b_j$ and $a_j$, we  clearly have
$[U_j(z_j),U_k(z_k)]=[V_j(z_j),V_k(z_k)]=0$, for $j\neq k$.

(2) Since the operators $U$ and $V$ are, for generic $z_j$, unbounded, definition (\ref{32}) makes sense only if
$\varphi_{\bf 0}\in D(U)$ and $\Psi_{\bf 0}\in D(V)$, a condition which
will be assumed here. In \cite{bagpb1} it was proven that, for
instance, this is granted when $\F_\varphi$ and $\F_\Psi$ are Riesz
bases.

(3) The set $\D$ could,  in
principle, be a proper subset of $\Bbb{C}$.

\vspace{2mm}

It is possible to write the vectors $\varphi(z_1,z_2)$ and
$\Psi(z_1,z_2)$ in terms of the vectors of $\F_\Psi$ and
$\F_\varphi$ as \be
\left\{
\begin{array}{ll}
\varphi(z_1,z_2,\ldots,z_d)=e^{-(|z_1|^2+|z_2|^2+\ldots+|z_d|^2)/2}\,\sum_{\bf n}\,\frac{z_1^{n_1}\,z_2^{n_2}\cdots z_d^{n_d}}{\sqrt{n_1!\,n_2!\ldots n_d!}}\,\varphi_{\bf n},\\
\\
\Psi(z_1,z_2,\ldots,z_d)=e^{-(|z_1|^2+|z_2|^2+\ldots+|z_d|^2)/2}\,\sum_{\bf n}\,\frac{z_1^{n_1}\,z_2^{n_2}\cdots z_d^{n_d}}{\sqrt{n_1!\,n_2!\ldots n_d!}}\,\Psi_{\bf n}.\end{array}
\right.
\label{33x}\en

These vectors are called {\em coherent} since they are eigenstates
of the lowering operators. Indeed we can check that \be
a_j\varphi(z_1,z_2,\ldots,z_d)=z_j\varphi(z_1,z_2,\ldots,z_d), \qquad
b_j^\dagger\Psi(z_1,z_2,\ldots,z_d)=z_j\Psi(z_1,z_2,\ldots,z_d), \label{34x}\en for
$j=1,2,\ldots,d$ and $z_j\in\D$. It is also a standard exercise, putting
$z_j=r_j\,e^{i\theta_j}$, to check that the following operator
equalities hold: \be
\left\{
\begin{array}{ll}
\frac{1}{\pi^d}\int_{\Bbb{C}}\,dz_1\int_{\Bbb{C}}\,dz_2\,\ldots\int_{\Bbb{C}}\,dz_d\,
|\varphi(z_1,z_2,\ldots,z_d)><\varphi(z_1,z_2,\ldots,z_d)|=S_\varphi, \\
\frac{1}{\pi^d}\int_{\Bbb{C}}\,dz_1\int_{\Bbb{C}}\,dz_2\,\ldots\int_{\Bbb{C}}\,dz_d\,
|\Psi(z_1,z_2,\ldots,z_d)><\Psi(z_1,z_2,\ldots,z_d)|=S_\Psi,\end{array}
\right. \label{35a}\en as well as
\be \frac{1}{\pi^d}\int_{\Bbb{C}}\,dz_1\int_{\Bbb{C}}\,dz_2\,\ldots\int_{\Bbb{C}}\,dz_d\,
|\varphi(z_1,z_2,\ldots,z_d)><\Psi(z_1,z_2,\ldots,z_d)|=\1, \label{36a}\en which are
written in convenient bra-ket notation. It should be said that
these equalities are, most of the times, only formal results.
Indeed, extending our result in \cite{abg}, we can prove the following

\vspace{2mm}

\begin{thm}\label{thm1} Let $a_j$, $b_j$, $\F_\varphi$, $\F_\Psi$, $\varphi(z_1,z_2,\ldots,z_d)$ and $\Psi(z_1,z_2,\ldots,z_d)$ be as above. Let us assume that
(1) $\F_\varphi$, $\F_\Psi$ are Riesz bases; (2) $\F_\varphi$,
$\F_\Psi$ are biorthogonal. Then (\ref{36a}) holds true.

\end{thm}

Suppose therefore that
the above construction gives coherent states that do not satisfy a
resolution of the identity. Then, since $\F_\varphi$ and
$\F_\Psi$ are automatically biorthogonal, they cannot be Riesz
bases. An example will be discussed in Section V.

\subsection{Relations with ordinary bosons}
In a recent paper, \cite{bagpb6}, we have considered the relations between PB, RPB and ordinary bosons. The role of unbounded operators appears to be crucial, and many domain problems have to be considered and solved. Here we just cite the two main theorems, referring to \cite{bagpb6} for the proofs and other details. We begin with the following theorem, concerning  RPB.

\begin{thm}
Let $a$ and $b$ be two operators on $\Hil$ satisfying  $[a,b]=\1$, and for which Assumptions 1, 2, 3 and 4 above are satisfied. Then an unbounded, densely defined, operator $c$ on $\Hil$ exists, together with a positive bounded operator $T$ with bounded inverse $T^{-1}$, such that $[c,c^\dagger]=\1$. Moreover
\be
a=TcT^{-1},\qquad b=Tc^\dagger T^{-1}.
\label{aa1}\en
Viceversa, given an unbounded, densely defined, operator $c$ on $\Hil$ satisfying $[c,c^\dagger]=\1$ and a positive bounded operator $T$ with bounded inverse $T^{-1}$, two operators $a$ and $b$ can be introduced as in (\ref{aa1}) for which $[a,b]=\1$ and Assumptions 1, 2, 3 and 4 above are satisfied.

\end{thm}

In \cite{bagpb6} we have also proven that, for ordinary pseudo bosons, the existence of a bounded $T$ with bounded inverse is not guaranteed at all! Indeed we have:

\begin{thm}
Let $a$ and $b$ be two operators on $\Hil$ satisfying  $[a,b]=\1$, and for which Assumptions 1, 2, and 3   are satisfied. Then two unbounded, densely defined, operators $c$ and $R$ on $\Hil$ exist,  such that $[c,c^\dagger]=\1$ and $R$ is positive, self adjoint and admits an unbounded inverse $R^{-1}$ . Moreover
\be
a=RcR^{-1},\qquad b=Rc^\dagger R^{-1},
\label{aa2}\en
and, introducing $\hat\varphi_n=\frac{{c^\dagger}^n}{\sqrt{n!}}\,\hat\varphi_0$, $c\varphi_0=0$, then $\hat\varphi_n\in D(R)\cap D(R^{-1})$ for all $n\geq0$. Also, the sets $\{R\hat\varphi_n\}$ and $\{R^{-1}\hat\varphi_n\}$ are biorthogonal bases of $\Hil$.

Viceversa, let us consider two unbounded, densely defined, operators $c$ and $R$ on $\Hil$ satisfying $[c,c^\dagger]=\1$ with $R$ positive, self-adjoint  with unbounded inverse $R^{-1}$. Suppose that, introducing $\hat\varphi_n$ as above, $\hat\varphi_n\in D(R)\cap D(R^{-1})$ for all $n\geq0$ and that the sets $\{R\hat\varphi_n\}$ and $\{R^{-1}\hat\varphi_n\}$ are biorthogonal bases of $\Hil$. Then  two operators $a$ and $b$ can be introduced for which $[a,b]=\1$, and for which equations (\ref{aa2}) and Assumptions 1, 2, and 3  are satisfied.

\end{thm}

It is clear the difference between the two situations. As we will see in the rest of the paper, this difference is important since {\em physical} examples seem to be related to PB rather than to RPB, so that all these subtleties on domains of operators turn out to be essential.

\section{Examples from mathematics}

We begin our review with two examples of RPB in $d=1$, mainly mathematically motivated. More details can be found in \cite{bagcal}.

\subsection{An example in coordinate space}

Let $\rho(x)$ be a measurable complex valued function satisfying, almost
everywhere (a.e.) on $\Bbb R$, the inequality $\alpha \leq \left\vert
\rho \left( x\right) \right\vert \leq \beta$, for some
$\alpha$ and $\beta$ with $0<\alpha\leq\beta<\infty$. Hence $\rho(x)$ is invertible and both $\rho(x)$ and
$\rho ^{-1}(x)$ belong to $\Lc^\infty(\mathbb{R})$, with $\beta^{-1}
\leq \left\vert\, \rho \left( x\right)^{-1} \right\vert \leq
\alpha^{-1}$. Let us now define a multiplication operator $X$ acting
on $\Lc^2(\Bbb R)$ as follows: $X\,f(x)=\rho(x)\,f(x)$, for all
$f(x)\in  \Lc^2(\Bbb R)$. This operator is bounded and admits a bounded inverse $X^{-1}$,
 $X^{-1}\,f(x)=\rho(x)^{-1}\,f(x)$ for all $f(x)\in
\Lc^2(\Bbb R)$. $X$ can therefore be used to build up a Riesz basis
for $\Lc^2(\Bbb R)$. For that we consider an o.n. basis
$\E=\{\hat{e}_{n}\left( x\right) \in
\mathcal{L}^{2}\left(\mathbb{R}\right) ,\,\
n\in{\Bbb{N}}_0:=\mathbb{N}\cup\{0\}\}$. Then $\F_\varphi=
\{\varphi _{n}(x)=X\hat{e}_{n}\left( x\right) =\rho \left( x\right)
\hat{e}_{n}\left( x\right)\left( x\right) \in
\Lc^{2}\left(\mathbb{R}\right) ,\,n\in{\Bbb{N}}_0\}$ is such a
basis. The vectors $\varphi_n(x)$ are not normalized, in general, and satisfy the following inequality:
$\alpha\leq  \|\varphi_n\|\leq
\beta$, for all $n\in{\Bbb{N}}_0$.

The operator $S_\varphi=\sum_{n=0}^\infty\,|\varphi_n><\varphi_n|$
can be easily computed: since $|\varphi_n>=X|\hat e_n>$,
$<\varphi_n|=<\hat e_n|X^\dagger$ and $\sum_{n=0}^\infty\,|\hat
e_n><\hat e_n|=\1$,  we get $S_\varphi=X\,X^\dagger$. Hence
$(S_\varphi\,f)(x)=|\rho(x)|^2\,f(x)$, for all $f(x)\in  \Lc^2(\Bbb R)$.
This operator is obviously bounded with bounded inverse
$S_\varphi^{-1}=\left(X^{\dagger}\right)^{-1}X^{-1}$, and is self-adjoint.

The dual frame of $\F_\varphi$ is $$\F_\Psi=  \left\{\Psi
_{n}(x)=S_\varphi^{-1}\varphi_n(x)=\frac{\varphi_n(x)}{|\rho(x)|^2}=\frac{\rho(x)}{|\rho(x)|^2}\,
 \hat{e}_{n}(x) \in \Lc^{2}\left(\mathbb{R}\right) ,\,n\in{\Bbb{N}}_0\right\},$$
which again consists of not normalized functions. It
is evident that $\F_\varphi$ and $\F_\Psi$ are biorthogonal and that
they produce the resolutions of the identity $
\sum_{n=0}^\infty |\varphi_n><\Psi_n|=\sum_{n=0}^\infty
|\Psi_n><\varphi_n|=\sum_{n=0}^\infty |\hat e_n><\hat e_n|=\1$. The next ingredient of our construction is
the o.n. basis $\F_{\hat\varphi}$ which is obtained as
$\F_{\hat\varphi}:=\{\hat\varphi_n(x):=S_\varphi^{-1/2}\varphi_n(x)=\frac{\varphi_n(x)}{|\rho(x)|}=\frac{\rho(x)}{|\rho(x)|}\,
 \hat{e}_{n}(x),\,n\geq0\}$. Incidentally we observe that, for real $\rho(x)$, this set coincides with $\E$.
 Also, if $\rho(x)=1$ a.e. in $\Bbb R$ all the sets collapse in the original one, $\E$. We are now ready to define a raising
 operator  $a_\varphi^\dagger$ on  $\F_{\hat\varphi}$ by $a_\varphi^\dagger\hat\varphi_n=\sqrt{n+1}\,\hat\varphi_{n+1}$, $n\geq0$.
 Hence the associated lowering operator $a_\varphi$ is naturally defined as  $a_\varphi\hat\varphi_n=\sqrt{n}\,\hat\varphi_{n-1}$,
 $n\geq0$, and $[a_\varphi,a_\varphi^\dagger]=\1$. The  pseudo-bosonic operators $a$ and $b$ are defined as $a:=S_\varphi^{1/2}\,a_\varphi\,S_\varphi^{-1/2}$
 and $b:=S_\varphi^{1/2}\,a_\varphi^\dagger\,S_\varphi^{-1/2}$. Their action on a given function $f(x)\in\Lc^2(\Bbb R)$ can be deduced to be
 $$
 (a\,f)(x)=|\rho(x)|\,\sum_{n=1}^\infty\,\sqrt{n} c_n \hat \varphi_{n-1}(x), \quad
 (b\,f)(x)=|\rho(x)|\,\sum_{n=0}^\infty\,\sqrt{n+1} c_n \hat \varphi_{n+1}(x),
 $$
 where the coefficients $c_n$ are defined as follows: $c_n=\int_{\Bbb R}\overline{\hat{\varphi}_{n}(x') }\,\frac{f(
x') }{\left\vert \rho \left( x'\right) \right\vert }\,dx'$.
Analogously we can find the action of $a^\dagger$ and $b^\dagger$ on
a generic $f(x)\in\Lc^2(\Bbb R)$. We get
 $$
 (a^\dagger\,f)(x)=|\rho(x)|^{-1}\,\sum_{n=0}^\infty\,\sqrt{n+1}\, d_n \hat \varphi_{n+1}(x),
 $$
 and
 $$
 (b^\dagger\,f)(x)=|\rho(x)|^{-1}\,\sum_{n=1}^\infty\,\sqrt{n}\, d_n \hat \varphi_{n-1}(x),
 $$
 where  $d_n=\int_{\Bbb R}\overline{\hat{\varphi}_{n}(x') }\,f(
x')\left\vert \rho( x') \right\vert \,dx'$. From these formulas we
deduce that, but for the trivial situation when $\rho(x)=1$ a.e.,
$a^\dagger\neq b$. The same formulas, taking $f=\varphi_n$, also show that
$(a\,\varphi_n)(x)=\sqrt{n}\,\,\varphi_{n-1}(x)$ and
$(b\,\varphi_n)(x)=\sqrt{n+1}\,\,\varphi_{n+1}(x)$: $a$ and $b$ are respectively a lowering and a raising
operator for $\F_\varphi$.  Moreover, the operator $a^\dagger$ is in fact a
raising operator and $b^\dagger$ a lowering operator, {\bf but with
respect to the dual basis} $\F_\Psi$: $\left( a^{\dagger}\Psi
_{m}\right) ( x) = \sqrt{m+1}\,\Psi _{m+1}( x)$ and $\left( b^{\dag
}\Psi _{m}\right) \left( x\right)  =\sqrt{ m}\,\Psi _{m-1}\left(
x\right)$. The pseudo-bosonic commutation rules
$[a,b]=\1$ are now easily recovered, as well as the eigenvalue
equations $N\varphi_n=n\varphi_n$ and
$\N\Psi_n=n\Psi_n$, $n\geq0$. Yet, in Dirac's bra-ket language, $a$ and $b$ can be written as
$$
a=\sum_{n=1}^\infty\,\sqrt{n}\,|\varphi_{n-1}\left>\right<\Psi_n|, \qquad b=\sum_{n=0}^\infty\,\sqrt{n+1}\,|\varphi_{n+1}\left>\right<\Psi_n|.
$$
We can now introduce three
different classes of (bi)-coherent states. The first  arises from the
o.n. basis $\F_{\hat\varphi}$:
$\hat\varphi(z;x)=e^{-|z|^2/2}\,\sum_{n=0}^\infty\,\frac{z^n}{\sqrt{n!}}\,\hat\varphi_n(x)$.
These are standard coherent states, clearly: they are normalized for
all $z\in\Bbb{C}$, satisfy a resolution of the identity which we
write here as
$$
f(x)=\frac{1}{\pi}\int_{\Bbb{C}}\,K(z)\,\hat\varphi(z;x)\,dz, \mbox{ where } K(z)=\int_{\Bbb{R}} \overline{\hat\varphi(z;x')}\,f(x')\,dx',
$$
for all $f(x)\in\Lc^2(\Bbb{R})$. Moreover we have
$a_\varphi\hat\varphi(z;x)=z\,\hat\varphi(z;x)$ for all $x\in\Bbb{R}$. They
finally saturate the Heisenberg uncertainty relation.

We can further define
$\varphi(z;x):=S_\varphi^{1/2}\hat\varphi(z;x)=e^{-|z|^2/2}\,\sum_{n=0}^\infty\,\frac{z^n}{\sqrt{n!}}\,\varphi_n(x)$
and their dual
$\Psi(z;x):=S_\varphi^{-1/2}\hat\varphi(z;x)=e^{-|z|^2/2}\,\sum_{n=0}^\infty\,\frac{z^n}{\sqrt{n!}}\,\Psi_n(x)$.
For these two states we can check explicitly (\ref{35a})-(\ref{36a}).

\subsection{An example in momentum space}

Let now $\tilde\alpha(p)$ be a measurable function in $\Lc^\infty(\Bbb R)$ with inverse also in $\Lc^\infty(\Bbb R)$: $\alpha \leq \left\vert \tilde\alpha(p) \right\vert \leq \beta$, for some positive $\alpha$ and $\beta$
 with $0<\alpha\leq\beta<\infty$. Hence  $\beta^{-1} \leq \left\vert\, \tilde\alpha^{-1}(p) \right\vert
 \leq \alpha^{-1}$. Let us now define a multiplication operator $X$ acting on $\Lc^2(\Bbb R)$ as follows: $X\,\tilde f(p)=\tilde\alpha(p)
 \,\tilde f(p)$, for all $f(x)\in  \Lc^2(\Bbb R)$, and a second operator $T$ which acts on such a $f(x)$ as $(Tf)(x)=\left(F^{-1}XFf\right)(x)=F^{-1}
 \left(\tilde\alpha(p)\tilde f(p)\right)(x)$. Here $F$ and $F^{-1}$ are the Fourier transform and the inverse Fourier transform. $T$ is bounded,
 and admits a bounded inverse $T^{-1}=F^{-1}X^{-1}F$ which acts on $\Lc^2(\Bbb R)$
  as $(T^{-1}\,f)(x)=F^{-1}\left(\frac{\tilde f}{\tilde\alpha}\right)(x)$, for all $f(x)\in  \Lc^2(\Bbb
  R)$.

Let us now  consider the same o.n. basis $\E$ introduced before,
and let $\F_\varphi= \{\varphi
_{n}(x)=T\hat{e}_{n}(x)=F^{-1}(\tilde\alpha\,\tilde e_n)(x) \in
\Lc^{2}\left(\mathbb{R}\right) ,\,n\in{\Bbb{N}}_0\}$. Of course the
set $\tilde\E=\{\tilde e_n(p)=(Fe_n)(p),\,n\in{\Bbb{N}}_0\}$ is also
an o.n. basis of $\Lc^2(\Bbb{R})$.

The operator
$S_\varphi=\sum_{n=0}^\infty\,|\varphi_n><\varphi_n|$ is
$S_\varphi=T\,T^\dagger=(F^{-1}X)(F^{-1}X)^\dagger$. Its action on a given
$f(x)\in \Lc^2(\Bbb{R})$ is
$(S_\varphi f)(x)=F^{-1}\left(|\tilde\alpha|^2\tilde f\right)(x)$. $S_\varphi$
is clearly self-adjoint, bounded and invertible. Indeed we have
$\alpha^2\1\leq S_\varphi\leq \beta^2\1$, $S_\varphi^{-1}=F^{-1}(XX^\dagger)^{-1}F$,
$\beta^{-2}\1\leq S_\varphi^{-1}\leq \alpha^{-2}\1$.
Moreover we have $\F_\Psi= \{\Psi
_{n}(x)=(S_\varphi^{-1}\varphi_n)(x)=F^{-1}\left(\frac{\tilde\alpha}{|\tilde\alpha|^2}\,\tilde
e_n\right)(x),\,n\in{\Bbb{N}}_0\}$.

As in the previous example, the vectors $\varphi_n(x)$ and $\Psi_n(x)$ are not normalized, in
general.  Needless to say, both $\F_\varphi$ and $\F_\Psi$
are complete in $\Lc^2(\Bbb{R})$, since they are both Riesz bases by construction.
Moreover,  for all $f, g\in\Lc^2(\Bbb{R})$ we have
$$
\sum_{n=0}^\infty \left<f,\varphi_n\right>\left<\Psi_n,g\right>=\sum_{n=0}^\infty \left<\tilde f,\tilde\alpha\,\tilde e_n\right>
\left<\frac{\tilde\alpha}{|\tilde\alpha|^2}\,\tilde e_n,\tilde g\right>=\left<\tilde f,\tilde g\right>=\left< f, g\right>.
$$
Here we have used the resolution of the identity associated to
$\tilde\E$ and the isometry of $F$. Hence $\sum_{n=0}^\infty
\left|\varphi_n\right>\left<\Psi_n\right|=\1$. Analogously we can
check that $\sum_{n=0}^\infty
\left|\Psi_n\right>\left<\varphi_n\right|=\1$, and that
$\left<\Psi_n,\varphi_m\right>=\delta_{n,m}$: $\F_\varphi$ and
$\F_\Psi$ are biorthogonal and produce two (related) resolutions of
the identity.

The o.n. basis $\F_{\hat\varphi}$  is
$\F_{\hat\varphi}:=\{\hat\varphi_n(x):=S_\varphi^{-1/2}\varphi_n(x)=F^{-1}\left(\frac{\tilde\alpha}{|\tilde\alpha|}\,\tilde
e_n\right)(x),\,n\geq0\}$, which again coincides with $\E$ if
$\tilde\alpha(p)$ is taken real.  The raising and
lowering
 operators  on  $\F_{\hat\varphi}$ are introduced as usual: $a_\varphi^\dagger\hat\varphi_n=\sqrt{n+1}\,\hat\varphi_{n+1}$,
 $n\geq0$,
 and  $a_\varphi\hat\varphi_n=\sqrt{n}\,\hat\varphi_{n-1}$,
 $n\geq0$. Then $[a_\varphi,a_\varphi^\dagger]=\1$. The  pseudo-bosonic operators $a$ and $b$ are  $a:=S_\varphi^{1/2}\,a_\varphi\,S_\varphi^{-1/2}$
 and $b:=S_\varphi^{1/2}\,a_\varphi^\dagger\,S_\varphi^{-1/2}$. Their action on a given function $f(x)\in\Lc^2(\Bbb R)$
 is analogous to the one we have deduced in the previous example:
 $$
 (a\,f)(x)=\sum_{n=1}^\infty\,\sqrt{n}\, c_n F^{-1}\left(|\tilde\alpha| \tilde{\hat\varphi}_{n-1}\right)(x),
 $$
 and
 $$
 (b\,f)(x)=\sum_{n=0}^\infty\,\sqrt{n+1}\, c_n F^{-1}\left(|\tilde\alpha| \tilde{\hat\varphi}_{n+1}\right)(x),
 $$
 where the coefficients $c_n$ can be written as  $c_n=\left<\frac{1}{|\tilde\alpha|}\,\tilde{\hat\varphi}_{n}, \tilde
 f\right>$. The adjoints of these operators can be easily deduced, \cite{bagcal}, and we see that
$(a\,\varphi_n)(x)=\sqrt{n}\,\,\varphi_{n-1}(x)$,
$(b\,\varphi_n)(x)=\sqrt{n+1}\,\,\varphi_{n+1}(x)$,  and that
$\left( a^{\dagger}\Psi _{m}\right) ( x) = \sqrt{m+1}\,\Psi _{m+1}(
x)$ and $\left( b^{\dag }\Psi _{m}\right) \left( x\right)  =\sqrt{
m}\,\Psi _{m-1}\left( x\right)$: $a$ and $b$ are lowering and
raising operators for $\F_\varphi$,  while $b^\dagger$ and
$a^\dagger$ are lowering and raising operators for $\F_\Psi$.

The pseudo-bosonic commutation rules $[a,b]=\1$ and the eigenvalue
equations  $N\varphi_n=n\varphi_n$ and $\N\Psi_n=n\Psi_n$, $n\geq0$
are easily recovered. Moreover, as in the previous example,   we can
introduce the standard coherent states
$\hat\varphi(z;x)=e^{-|z|^2/2}\,\sum_{n=0}^\infty\,\frac{z^n}{\sqrt{n!}}\,\hat\varphi_n(x)$
and the bi-coherent states
$$\varphi(z;x):=S_\varphi^{1/2}\hat\varphi(z;x)=e^{-|z|^2/2}\,\sum_{n=0}^\infty\,\frac{z^n}{\sqrt{n!}}\,\varphi_n(x)$$
and their dual
$$\Psi(z;x):=S_\varphi^{-1/2}\hat\varphi(z;x)=e^{-|z|^2/2}\,\sum_{n=0}^\infty\,\frac{z^n}{\sqrt{n!}}\,\Psi_n(x).$$
For these two states we can check explicitly the equalities in (\ref{35a})-(\ref{36a}).

\section{Examples from Susy}

In this section we discuss an example physically motivated in $d=1$. More details can be found in \cite{bagpb2}. We take  $\Hil=\Lc^2(\Bbb R)$ and we look for solutions of the commutation rule in (\ref{21}) of the following form:
\be
a=\frac{1}{\sqrt{2}}\left(\frac{d}{dx}+W_a(x)\right), \qquad b=\frac{1}{\sqrt{2}}\left(-\frac{d}{dx}+W_b(x)\right).
\label{31}\en
Here $W_j(x)$, $j=a,b$, are two functions such that $W_a(x)\neq \overline{W_b(x)}$. Hence $b^\dagger\neq a$. For future convenience we will assume that both $W_a(x)$ and $W_b(x)$ are sufficiently regular functions, for example that they are differentiable. We will show how to fix these functions in such a way Assumptions 1-4  are satisfied, while explicit choices of $W_a(x)$ and $W_b(x)$ will be considered later on. The starting point is that $a$ and $b$ are required to satisfy $[a,b]=\1$. A straightforward computation implies that $W_a(x)$ and $W_b(x)$ must obey the following simple equality:
\be
W_a(x)+W_b(x)=2x+\alpha,
\label{32b}\en
where $\alpha$ is an arbitrary complex integration constant.  In particular, if we compute $N=ba$ and we use (\ref{32b}) we get
\be
N=b\,a=\frac{1}{2}\left(-\frac{d^2}{dx^2}+U(x)\,\frac{d}{dx}+V(x)\right),
\label{33}\en
where $V(x):=W_a(x)(2x+\alpha-W_a(x))-W_a'(x)$ and $U(x):=2x+\alpha-2 W_a(x)$.

We observe that the approach we are adopting here is just an extension of the standard ideas of SUSY quantum mechanics, where the operator $N$ is just the hamiltonian of the system and $W(x)=W_a(x)=W_b(x)$ is the so-called super-potential, which is related to the (physical) potential via a Riccati equation. For this reason we still call both $W_a(x)$ and $W_b(x)$ superpotentials. In the first part of this section we will limit ourselves to real functions $W_a(x)$ and $W_b(x)$, extending our results to complex superpotentials in the second part. This will produce some interesting results, as we will see. Hence $\alpha$ in (\ref{32b}) will be taken to be real, for the moment.

\vspace{2mm}

{\bf Remark:--} It may be interesting to observe that, if $U(x)\equiv0$, then $N$ in (\ref{33}) looks like a one-dimensional hamiltonian (at least formally: we should  check for self-adjointness of the operator). This choice produces a  well known situation: $U(x)=0$ implies that $W_a(x)=x+\frac{\alpha}{2}$ and $V(x)=\left(x+\frac{\alpha}{2}\right)^2-1$ so that $N$ becomes, but for an unessential constant, the hamiltonian of a shifted harmonic oscillator,  $N=\frac{1}{2}\left(-\frac{d^2}{dx^2}+\left(x+\frac{\alpha}{2}\right)^2-1\right)$. This is in agreement with the fact that $W_b(x)=2x+\alpha-W_a(x)=W_a(x)$. Hence, if $\alpha$ is real, we deduce that $a^\dagger=b$ and we recover the ordinary CCR.

\vspace{2mm}

The next step consists in solving the two equations $a\varphi_0(x)=0$ and $b^\dagger\Psi_0(x)=0$, looking for solutions in $\Hil=\Lc^2(\Bbb R)$. These solutions are easily found: \be\varphi_0(x)=N_\varphi\exp\{-w_a(x)\},\qquad \Psi_0(x)=N_\Psi\exp\{-w_b(x)\},\label{34}\en where $N_\varphi$ and $N_\Psi$ are normalization constants which can be written as $N_\varphi=\varphi_0(0)\,\exp\{w_a(0)\}$ and $N_\Psi=\Psi_0(0)\,\exp\{w_b(0)\}$. We have introduced  the following functions
\be
w_j(x)=\int W_j(x)\,dx,
\label{35}\en
$j=a,b$. Of course since $\varphi_0(x)$ and $\Psi_0(x)$ must be square integrable, this imposes some constraints on the asymptotic behaviors of the $w_j(x)$'s and, as a consequence,  on the $W_j(x)$'s. We will consider this aspect in more details below.

It is possible to prove, \cite{bagpb2}, that, independently of the analytic expressions of the $w_j(x)$'s, the following is true:
\be
\varphi_n(x)=N_n^\varphi\,p_n(x)\,\exp\{-w_a(x)\},\qquad N_n^\varphi=\frac{\varphi_0(0)\,\exp\{w_a(0)\}}{\sqrt{n!\,2^n}},
\label{36b}\en
and
\be
\Psi_n(x)=N_n^\Psi\,p_n(x)\,\exp\{-w_b(x)\},\qquad N_n^\Psi=\frac{\Psi_0(0)\,\exp\{w_b(0)\}}{\sqrt{n!\,2^n}},
\label{37b}\en
where an {\bf unique} polynomial $p_n(x)$ appears both in $\varphi_n(x)$ and in $\Psi_n(x)$. This is defined recursively as follows: $p_0(x)=1$ and $p_{n+1}(x)=(2x+\alpha)p_n(x)-p_n'(x)$, $n\geq 0$. Therefore $p_1(x)=2x+\alpha$, $p_2(x)=(2x+\alpha)^2-2$, $p_3(x)=(2x+\alpha)\left((2x+\alpha)^2-6\right)$ and so on. The proof of this claim is based on induction. Hence, if both $w_a(x)$ and $w_b(x)$ diverges to $+\infty$ when $|x|\rightarrow\infty$ at least as $|x|^\mu$ for some positive $\mu$, Assumptions 1 and 2 are satisfied.

Let us now observe that, because of (\ref{32b}), we also have that
\be
w_a(x)+w_b(x)=x^2+\alpha x+\beta,
\label{38b}\en
where $\beta$ is a second integration constant which again we take real for the moment. Therefore, since  $w_a(x)$ and $w_b(x)$ should diverge to $+\infty$ for large $|x|$ as $|x|^{\mu_j}$ for some $\mu_j>0$, $j=a,b$, this equality also fixes an upper bound for the $\mu_j$'s: we must have $0<\mu_j\leq2$, $j=a,b$.

Obviously we have
\be
N\varphi_n(x)=n\,\varphi_n(x), \qquad N^\dagger \Psi_n(x)=n\Psi_n(x),
\label{39y}\en
for all $n\geq0$. Moreover these functions are biorthogonal:
\be
\left<\varphi_n,\Psi_m\right>=\delta_{n,m}\left<\varphi_0,\Psi_0\right>,
\label{310b1}\en
i.e.,
\be
\sqrt{\frac{1}{n!m!2^{n+m}}}\int_{\Bbb{R}}p_n(x)p_m(x)e^{-(x^2+\alpha x+\beta)}\,dx=\delta_{n,m}\int_{\Bbb{R}}e^{-(x^2+\alpha x+\beta)}\,dx=
\delta_{n,m}\,\sqrt{\pi}\,e^{\alpha^2/4-\beta}
\label{311}\en

\vspace{3mm}

{\bf Remark:--} Our $p_n(x)$ are  related to Hermite polynomials since we can check that $p_n(x)=(-1)^ne^{x^2+\alpha x}\,\frac{d^n}{dx^n}\,e^{-(x^2+\alpha x)}$, for all $n\geq 0$.

\vspace{2mm}

We are now ready to check if or when Assumption 3 is verified, that is whether   $\Hil_\varphi=\Hil_\Psi=\Hil$.

To check this we first observe that $\F_\varphi$ is complete in $\Hil$ if and only if the set $\F_\pi^{(a)}=\left\{\pi_n^{(a)}(x):=x^n\,e^{-w_a(x)},\,n\geq0\right\}$ is complete in $\Hil$. Analogously, $\F_\Psi$ is complete in $\Hil$ if and only if the set $\F_\pi^{(b)}=\left\{\pi_n^{(b)}(x):=x^n\,e^{-w_b(x)},\,n\geq0\right\}$ is complete in $\Hil$. But, \cite{kolfom}, if $\rho(x)$ is a Lebesgue-measurable function which is different from zero a.e. in $\Bbb R$ and if there exist two positive constants $\delta, C$ such that $|\rho(x)|\leq C\,e^{-\delta|x|}$ a.e. in $\Bbb R$, then the set $\left\{x^n\,\rho(x)\right\}$ is complete in $\Lc^2(\Bbb{R})$.

This suggests to consider the following constraint on the asymptotic behavior of the $w_j(x)$'s: for Assumption 3 to be satisfied it is sufficient that four positive constants $C_j,\,\delta_j$, $j=a,b$ exist such that
\be
\left|e^{-w_j(x)}\right|\leq C_j\,e^{-\delta_j|x|},
\label{311b}
 \en
 $j=a,b$, holds a.e. in $\Bbb R$. It should be noticed that this condition is stronger than the one required for Assumptions 1 and 2 to hold, since for instance it is not satisfied if $w_a(x)\simeq|x|^{1/2}$ for large $|x|$.

Using now the biorthogonality of the sets $\F_\varphi$ and $\F_\Psi$, and their completeness in $\Lc^2(\Bbb R)$ we can write
\be
\frac{1}{\left<\Psi_0,\varphi_0\right>}\,\sum_{k=0}^\infty
|\varphi_k><\Psi_k|=\frac{1}{\left<\varphi_0,\Psi_0\right>}\,\sum_{k=0}^\infty
|\Psi_k><\varphi_k|=\1,
\label{313}\en
where $\1$ is the identity operator on $\Lc^2(\Bbb R)$ and the overall constants $\left<\Psi_0,\varphi_0\right>^{-1}$ and $\left<\varphi_0,\Psi_0\right>^{-1}$ appear because of (\ref{310b1}).

Suppose now that we are interested in going from $\F_\varphi$ to $\F_\Psi$ and viceversa. In other words we are now interested to introduce an invertible operator $S$ mapping each $\varphi_n$ into $\Psi_n$, $S\varphi_n=\Psi_n$, whose inverse of course satisfies $S^{-1}\Psi_n=\varphi_n$, for all $n\geq 0$. As we have already seen in Section II, such an operator in general can be defined but might be unbounded, so a special care is required. In fact, it is clear that $S$ is nothing but the operator $S_\Psi$ introduced in Section II. A formal expansion of these operators is
\be
S_\Psi=\frac{1}{\left<\Psi_0,\varphi_0\right>}\,\sum_{k=0}^\infty
|\Psi_k><\Psi_k|,\qquad S_\Psi^{-1}=\frac{1}{\left<\varphi_0,\Psi_0\right>}\,\sum_{k=0}^\infty
|\varphi_k><\varphi_k|.
 \label{314}\en
It is quite easy to check that, again at least formally, $S_\Psi S_\Psi^{-1}=S_\Psi^{-1}S_\Psi=\1$. Due to the analytic expressions (\ref{36b}) and (\ref{37b}) of our wave-functions $\varphi_n(x)$ and $\Psi_n(x)$, we  deduce that
\be
S_\Psi=\frac{\Psi_0(0)}{\varphi_0(0)}\,\frac{e^{\delta w_a(x)}}{e^{\delta w_b(x)}},\qquad
S_\Psi^{-1}=\frac{\varphi_0(0)}{\Psi_0(0)}\,\frac{e^{\delta w_b(x)}}{e^{\delta w_a(x)}},
\label{315}\en
where we have introduced $\delta w_j(x):=w_j(x)-w_j(0)$, $j=a,b$. A sufficient condition for both $S_\Psi$ and $S_\Psi^{-1}$ to be bounded operators from $\Lc^2(\Bbb R)$ into itself is now easily deduced using  equation (\ref{38b}), which implies that $\frac{e^{\delta w_a(x)}}{e^{\delta w_b(x)}}=\frac{e^{2\delta w_a(x)}}{e^{x^2+\alpha x}}$ and $\frac{e^{\delta w_b(x)}}{e^{\delta w_a(x)}}=\frac{e^{x^2+\alpha x}}{e^{2\delta w_a(x)}}$:

{\em if $\frac{e^{2\delta w_a(x)}}{e^{x^2+\alpha x}}\in \Lc^\infty(\Bbb R)$, then $S_\Psi\in B(\Lc^2(\Bbb R))$. Moreover, if $\frac{e^{x^2+\alpha x}}{e^{2\delta w_a(x)}}\in \Lc^\infty(\Bbb R)$,  also $S_\Psi^{-1}\in B(\Lc^2(\Bbb R))$.}

It is clear that this boundedness assumption imposes further limitations on the functions $w_j(x)$'s and, as a consequence, on the $W_j(x)$'s.

\subsection{What if the superpotentials are complex?}

The above result on the boundedness of $S_\Psi$ and $S_\Psi^{-1}$ displays the relevance of $\alpha$: suppose $\alpha\neq0$. If $\delta w_a(x)$ behaves as $x^2/2$ for large $|x|$ then $\frac{e^{2\delta w_a(x)}}{e^{x^2+\alpha x}}$ and $\frac{e^{x^2+\alpha x}}{e^{2\delta w_a(x)}}$ cannot be bounded for both positive and negative $x$. This is not true if $\alpha$ is purely imaginary, of course: both these fractions are bounded functions so that $S_\Psi$ and $S_\Psi^{-1}$ are  bounded operators. That's why this choice is so interesting for us. In this case formulas (\ref{36b}) and (\ref{37b}) look like
\be
\varphi_n(x)=N_n^\varphi\,p_n(x)\,\exp\{-w_a(x)\},\qquad N_n^\varphi=\frac{\varphi_0(0)\,\exp\{w_a(0)\}}{\sqrt{n!\,2^n}},
\label{316}\en
and
\be
\Psi_n(x)=N_n^\Psi\,\overline{p_n(x)}\,\exp\{-\overline{w_b(x)}\},\qquad N_n^\Psi=\frac{\Psi_0(0)\,\exp\{\overline{w_b(0)}\}}{\sqrt{n!\,2^n}},
\label{317}\en
where $p_n(x)$ is defined as before.   Next we find that
\be
\left<\varphi_n,\Psi_m\right>=\delta_{n,m}\left<\varphi_0,\Psi_0\right>=\delta_{n,m}\,\sqrt{\pi}\,\Psi_0(0)\,\overline{\varphi_0(0)}\,e^{\overline{\alpha}^2/4}
\label{318}\en
The main difference arises in the analytic expression of $S_\Psi$ and of $S_\Psi^{-1}$. For that it is necessary to introduce the operator of complex conjugation $C$ which acts on a generic function $f(x)\in\Lc^2(\Bbb R)$ as follows: $Cf(x)=\overline{f(x)}$. $C$ is antilinear and idempotent: $C^2=\1$. Hence $C=C^{-1}$. While formulas  (\ref{314}) are still true,  (\ref{315}) must be replaced by
\be
S_\Psi=C\,\frac{\overline{\Psi_0(0)}}{\varphi_0(0)}\,\frac{e^{\delta w_a(x)}}{e^{\delta w_b(x)}},\qquad
S_\Psi^{-1}=\frac{\varphi_0(0)}{\overline{\Psi_0(0)}}\,\frac{e^{\delta w_b(x)}}{e^{\delta w_a(x)}}\,C,
\label{319}\en
It is a straightforward computation to check that they are indeed the inverse of one another and that $S_\Psi\varphi_n(x)=\Psi_n(x)$, $S_\Psi^{-1}\Psi_n(x)=\varphi_n(x)$ for all $n\geq0$. As for the norms of $S_\Psi$ and $S_\Psi^{-1}$, they are not affected by the presence of $C$ and of the complex conjugation in $\Psi_0(0)$: once again $S_\Psi$ and $S_\Psi^{-1}$ are bounded if both $\frac{e^{2\delta w_a(x)}}{e^{x^2+\alpha x}}$ and  $\frac{e^{x^2+\alpha x}}{e^{2\delta w_a(x)}}$ belong to $\Lc^\infty(\Bbb R)$. This means that, if $\alpha$ is purely imaginary, then both $S_\Psi$ and $S_\Psi^{-1}$ can be bounded and, as a consequence, $\F_\varphi$ and $\F_\Psi$ are Riesz bases: this is never possible if $\alpha$ is real.

\vspace{3mm}

Under Assumptions 1-4, some kind of {\em bi-coherent states} can be introduced, \cite{bagpb1}. Let us define the $z$-dependent operators
\be
U(z)=\exp\{z\,b-\overline{z}\,a\}, \qquad V(z)=\exp\{z\,a^\dagger-\overline{z}\,b^\dagger\},
\label{43}\en
$z\in\Bbb{C}$, and the following vectors:
\be
\varphi(z)=U(z)\varphi_0=e^{-|z|^2/2}\,\sum_{n=0}^\infty\,\frac{z^n}{\sqrt{n!}}\,\varphi_n,\qquad \Psi(z)=V(z)\,\Psi_0=e^{-|z|^2/2}\,\sum_{n=0}^\infty\,\frac{z^n}{\sqrt{n!}}\,\Psi_n.
\label{44}\en
Both these series are convergent for all possible $z\in\Bbb{C}$ due to the fact that $S_\Psi$ and $S_\Psi^{-1}$ are bounded, \cite{bagpb1}. These vectors are called (bi-){\em coherent} since they are eigenstates of our lowering operators. Indeed we can check that
\be
a\varphi(z)=z\varphi(z), \qquad b^\dagger\Psi(z)=z\Psi(z),
\label{45}\en
for all $z\in\Bbb{C}$. Moreover we have
\be
\frac{1}{\pi}\int_{\Bbb{C}}\,dz |\varphi(z)><\varphi(z)|=S_\Psi^{-1}, \qquad
\frac{1}{\pi}\int_{\Bbb{C}}\,dz |\Psi(z)><\Psi(z)|=S_\Psi,
\label{46y}\en
and
\be
\frac{1}{\pi}\int_{\Bbb{C}}\,dz |\varphi(z)><\Psi(z)|=
\frac{1}{\pi}\int_{\Bbb{C}}\,dz |\Psi(z)><\varphi(z)|=\1.
\label{47}\en
Of course, they can be associated to {\em standard} coherent states (i.e. coherent states built out of an o.n. basis) if $S_\Psi$ and $S_\Psi^{-1}$ are bounded, because of the properties of Riesz bases.

\subsection{Explicit examples}

We will now discuss two examples of our construction showing how easily Riesz bases can be constructed using a sort of perturbation technique applied to the harmonic oscillator.

\vspace{2mm}

{\bf Example 1:} we fix here $W_a(x)=x$. Hence $W_b(x)$ is fixed as in (\ref{32b}) just requiring that the related operators $a$ and $b$, see (\ref{31}), satisfy $[a,b]=\1$. Hence $W_b(x)=x+\alpha$ where, for the moment, we don't make any assumption on $\alpha$. Then we get $w_a(x)=\frac{x^2}{2}+k_a$ and $w_b(x)=\frac{x^2}{2}+\alpha x+k_b$. Here $k_a$ and $k_b$ are two integration constants which are, in general, complex. Their sum gives back $\beta$, see (\ref{38b}).

Using the inequality $e^{-x^2/2}\leq 2 e^{-|x|}$ it is  clear that $\left|e^{-w_a(x)}\right|\leq 2 \left|e^{-k_a}\right|\,e^{-|x|}$. Hence the set $\F_\varphi$ is a basis of $\Lc^2(\Bbb R)$. The same estimate, with $k_a$ replaced by $k_b$, can be repeated for $\left|e^{-w_b(x)}\right|$ if $\alpha$ is purely imaginary. If $\alpha$ is real this estimate does not work. However we get that $\left|e^{-w_b(x)}\right|\leq 2 \left|e^{-k_b}\right|\,e^{\alpha^2/2}\,e^{|\alpha|}\,e^{-|x|}$, which again implies that $\F_\Psi$ is a basis of $\Lc^2(\Bbb R)$.

A major difference arises if we require  these sets to be Riesz bases. Indeed, if $\alpha$ is purely imaginary,  $\left|\frac{e^{2\delta w_a(x)}}{e^{x^2+\alpha x}}\right|=\left|\frac{e^{x^2+\alpha x}}{e^{2\delta w_a(x)}}\right|=1$, so that both $S_\Psi$ and $S_\Psi^{-1}$ are bounded operators and $\F_\varphi$ and $\F_\Psi$ are automatically Riesz bases. If we rather look for real $\alpha$ such that the above fractions are both bounded functions, then the only possible choice is $\alpha=0$. Under this constraint we recover essentially the standard Hermite functions. This is not surprising since $\alpha=0$ implies $W_a(x)=W_b(x)$ and $a=b^\dagger$: we go back to the standard canonical commutation relation.

\vspace{2mm}

{\bf Example 2:} our above mentioned perturbation technique consists in adding a suitable perturbation to a {\em zero order} superpotential  $W_a^o(x)=x$. Let us consider a function $\Phi(x)$ which is differentiable and bounded in $\Bbb R$: $-\infty<\Phi_m\leq \Phi(x)\leq \Phi_M<\infty$, $\forall x\in \Bbb R$. Now we define $W_a(x)=x+\Phi'(x)$. Hence, by (\ref{32b}), $W_b(x)=x-\Phi'(x)+\alpha$. Consequently we have $w_a(x)=\frac{x^2}{2}+\Phi(x)+k_a$ and $w_b(x)=\frac{x^2}{2}-\Phi(x)+\alpha x+k_b$. The following inequalities hold:  $\left|e^{-w_a(x)}\right|\leq 2 \left|e^{-k_a}\right|\,e^{-\Phi_m}\,e^{-|x|}$ and  $\left|e^{-w_b(x)}\right|\leq 2 e^{\Phi_M} \left|e^{-k_b}\,e^{\alpha^2/2}\right|\,e^{|\alpha|}\,e^{-|x|}$. Therefore both $\F_\varphi$ and $\F_\Psi$ are bases for $\Lc^2(\Bbb R)$, independently of the nature of $\alpha$. As before, however, if $\alpha$ is purely imaginary then these are also Riesz bases, for the usual reason:  both $\left|\frac{e^{2\delta w_a(x)}}{e^{x^2+\alpha x}}\right|$ and $\left|\frac{e^{x^2+\alpha x}}{e^{2\delta w_a(x)}}\right|$ are bounded functions, as desired. The operators $a$ and $b$ in (\ref{31}) are
$$
a=\frac{1}{\sqrt{2}}\left(\frac{d}{dx}+x+\Phi'(x)\right), \qquad b=\frac{1}{\sqrt{2}}\left(-\frac{d}{dx}+x-\Phi'(x)+i\alpha_r\right),
$$
where $\alpha_r$ is an arbitrary but fixed real quantity.

A choice of $\Phi(x)$ which is not bounded but still {\em under control} is $\Phi(x)=\frac{\alpha x}{2}$. This produces $W_a(x)=W_b(x)=x+\frac{\alpha}{2}$, which is nothing but the shifted harmonic oscillator.

\vspace{2mm}
More details and examples can be found in \cite{bagpb2}.

\section{Example from non-hermitian quantum system}

The example which we consider here was motivated by the paper \cite{sun}, where the author consider a  simple modification of the CCR  in connection with non-hermitian quantum systems.  The starting point is a lowering operator $a$ acting on an Hilbert space $\Hil$ which, together with its adjoint $a^\dagger$, satisfies the CCR $[a,a^\dagger]=\1$. Then we consider the following simple deformation of $a$ and $a^\dagger$:
\be
A_\alpha=a-\alpha\,\1,\qquad B_\beta=a^\dagger-\beta\,\1.
\label{sun1}\en
It is clear that $[A_\alpha,B_\beta]=\1$ and that, if $\alpha\neq\overline\beta$, $A_\alpha\neq B_\beta^\dagger$. Our aim is to check that this example gives rise to PB which are not regular.

To check Assumption 1 first of all we have to find a vector $\varphi_0(\alpha)$ such that $A_\alpha\varphi_0(\alpha)=0$. Such a vector clearly exists since $A_\alpha\varphi_0(\alpha)=0$ can be written as $a\varphi_0(\alpha)=\alpha\varphi_0(\alpha)$. Hence it is enough to take $\varphi_0(\alpha)$ as the following coherent state:
$$
\varphi_0(\alpha)=U(\alpha)\varphi_0,
$$
where $U(\alpha)=e^{\alpha a^\dagger-\overline\alpha a}=e^{-|\alpha|^2/2}e^{\alpha a^\dagger} e^{\overline{\alpha} a}$ and $\varphi_0$ is the vacuum of $a$: $a\varphi_0=0$. Incidentally we recall that the set $\E=\{\varphi_n=\frac{(a^\dagger)^n}{\sqrt{n!}}\varphi_0,\,n\geq 0\}$ is an o.n. basis of $\Hil$. The fact that $\varphi_0(\alpha)$ belongs to $D^\infty(B_\beta)$ follows from the following estimate:  $$\|B_\beta^l\varphi_0(\alpha)\|\leq l! e^{|\overline\alpha-\beta|},$$ which holds for all $l\geq0$.

Let us now define a second vector $\Psi_0(\beta):=U(\overline\beta)\varphi_0$. This is a second coherent state, labeled by $\beta$, which satisfies Assumption 2: $B_\beta^\dagger\Psi_0(\beta)=0$ and $\Psi_0(\beta)\in D^\infty(A_\alpha^\dagger)$, since $\|(A_\alpha^\dagger)^l\Psi_0(\beta)\|\leq l! e^{|\overline\alpha-\beta|}$, for all $l\geq0$.

Now we introduce, following (\ref{22}), the vectors
\be
\varphi_n(\alpha,\beta):=\frac{B_\beta^n}{\sqrt{n!}}\,\varphi_0(\alpha), \qquad \Psi_n(\alpha,\beta):=\frac{(A_\alpha^\dagger)^n}{\sqrt{n!}}\,\Psi_0(\beta),
\label{61}\en
where the dependence on $\alpha$ and $\beta$ is written explicitly. It is possible to rewrite $\varphi_n(\alpha,\beta)$ and $\Psi_n(\alpha,\beta)$ in many different equivalent forms. For instance we have
\be
\varphi_n(\alpha,\beta)=V_\varphi(\alpha,\beta)\varphi_n, \qquad V_\varphi(\alpha,\beta)=e^{-|\alpha|^2/2}e^{\alpha a^\dagger}e^{-\beta a}=e^{\alpha(\beta-\overline\alpha)/2}e^{\alpha a^\dagger-\beta a}
\label{62}\en
and
\be
\Psi_n(\alpha,\beta)=V_\Psi(\alpha,\beta)\varphi_n, \qquad V_\Psi(\alpha,\beta)=e^{-|\beta|^2/2}e^{\overline\beta a^\dagger}e^{-\overline\alpha a}=e^{\overline\beta (\overline\alpha-\beta)/2}e^{\overline\beta  a^\dagger-\overline\alpha  a},
\label{63}\en
for all $n\geq0$. Notice that the operators $V_\varphi$ and $V_\Psi$,  are in general unbounded (see below) and densely defined  since each $\varphi_n$ belongs to $D(V_\varphi)$ and $D(V_\Psi)$.

\vspace{2mm}

{\bf Remark:--} It is interesting to notice that, if $\beta=\overline\alpha$, then everything collapses: $B_\beta^\dagger=A_\alpha$, $\varphi_0(\alpha)=\Psi_0(\beta)$, $\varphi_n(\alpha,\beta)=\Psi_n(\alpha,\beta)$ and, finally, $V_\varphi$ and $V_\Psi$ are unitary operators.\vspace{2mm}

Defining as usual $\F_\varphi^{(\alpha,\beta)}=\{\varphi_n(\alpha,\beta), n\geq0\}$ and $\F_\Psi^{(\alpha,\beta)}=\{\Psi_n(\alpha,\beta), n\geq0\}$, it is possible to check that both these sets are complete in $\Hil$: for that we rewrite $\varphi_n(\alpha,\beta)$ in the following equivalent way:
$$
\varphi_n(\alpha,\beta)=\frac{1}{\sqrt{n!}}\,e^{(\alpha\,\beta-\overline{\alpha}\,\overline{\beta})/2}\,
U(\overline{\beta}) (a^\dagger)^n U(\alpha-\overline{\beta})\varphi_0,
$$
and  use induction on $n$ and the properties of the unitary operators $U(\overline{\beta})$ and $U(\alpha-\overline{\beta})$. With the same techniques we can check that $\F_\Psi$ is complete in $\Hil$.

The vectors in $\F_\varphi$ and $\F_\Psi$ are also biorthogonal:
$$
\left<\varphi_n(\alpha,\beta),\Psi_m(\alpha,\beta)\right>=\delta_{n,m}
\exp\{\overline\alpha\overline\beta-\frac{1}{2}(|\alpha|^2+|\beta|^2)\},
$$
and biorthonormality could be recovered changing the normalization of $\varphi_0(\alpha,\beta)$ and $\Psi_0(\alpha,\beta)$.

As for Assumption 4, the situation is a bit more difficult:  if $\beta=\overline\alpha$, then both $\F_\varphi$ and $\F_\Psi$ became the same o.n. basis. However, whenever $\beta\neq\overline\alpha$, it is possible to prove that neither $\F_\varphi$ nor $\F_\Psi$ can be Riesz bases. Indeed, let us suppose, e.g., that $\F_\varphi$ is a Riesz basis. Then $\|\varphi_n(\alpha,\beta)\|$  must be uniformly bounded in $n$. On the other way, a direct estimates show that  $\|\varphi_n(\alpha,\beta)\|^2\geq 1+n|\overline\alpha-\beta|^2$, $\forall n \geq 0$, \cite{bagpb2}. Hence, uniform boundedness is compatible only with $\overline\alpha=\beta$, and we go back to ordinary bosons.
Moreover, since $\|V_\varphi(\alpha,\beta)\varphi_n\|^2=\|\varphi_n(\alpha,\beta)\|^2\geq 1+n|\overline\alpha-\beta|^2$, then $V_\varphi(\alpha,\beta)$ is, in general, unbounded, as already stated. Hence, $\F_\varphi$ and $\F_\Psi$ cannot be Riesz bases since, \cite{you},  two biorthogonal bases can be Riesz bases if and only if they are connected by a bounded operator with bounded inverse.

\subsection{Coherent states}

We now construct the coherent states associated to this model, working first in the coordinate representation. For that, calling $z=z_r+iz_i$, $z_r, z_i\in{\Bbb R}$, and $a=\frac{1}{\sqrt{2}}\left(x+\frac{d}{dx}\right)$,  the normalized solution of the eigenvalue equation $a\eta(x;z)=z\eta(x;z)$,  is, with a certain choice of phase in the normalization,
$\eta(x;z)=\frac{1}{\pi^{1/4}}\exp\left\{-\frac{x^2}{2}+\sqrt{2}\,z\,x-z_r^2\right\}$. Hence, calling $\Phi_\alpha(x;z)$ the eigenstate of $A_\alpha$ with eigenvalue $z$, $A_\alpha\Phi_\alpha(x;z)=z\Phi_\alpha(x;z)$, we get $\Phi_\alpha(x;z)=\eta(x;z+\alpha)$. Analogously, the eigenstate of $B_\beta^\dagger$ with eigenvalue $z$, $B_\beta^\dagger\Psi_\beta(x;z)=z\Psi_\beta(x;z)$, is  $\Psi_\beta(x;z)=\eta(x;z+\overline\beta)$. It is clear that $$\frac{1}{\pi}\int_{\Bbb{C}} \,dz\, |\Phi_\alpha(x;z)><\Phi_\alpha(x;z)|=\frac{1}{\pi}\int_{\Bbb{C}}\,dz\, |\Psi_\beta(x;z)><\Psi_\beta(x;z)|=\1.$$
On the other hand, taken $f,g\in\Hil$, we get
$$
\left<f,\left(\frac{1}{\pi}\int_{\Bbb{C}}\,dz\, |\Phi_\alpha(x;z)><\Psi_\beta(x;z)|\right)g\right>=
e^{-(\alpha_r-\beta_r)^2/2}\,\int_{\Bbb{R}}\,dx\,\overline{f(x)}\,g(x)e^{i\sqrt{2}\,(\alpha_i+\beta_i)},
$$
with obvious notation. Therefore, if $\alpha\neq\overline \beta$, the integral over $\Bbb C$ above does not produce the identity operator! To the same conclusion we arrive defining coherent states as in Section II.2. Following (\ref{32}) we  introduce
\be
\tilde U_{\alpha,\beta}(z)=\exp\left\{z\,B_\beta-\overline z A_\alpha\right\},\qquad \tilde V_{\alpha,\beta}(z)=\exp\left\{z\,A_\alpha^\dagger-\overline z B_\beta^\dagger\right\},
\label{65}\en
and two associated vectors
$$
\tilde\varphi_{\alpha,\beta}(z)=\tilde U_{\alpha,\beta}(z)\varphi_0, \qquad \tilde\Psi_{\alpha,\beta}(z)=\tilde V_{\alpha,\beta}(z)\varphi_0.
$$
They satisfy $A_\alpha\tilde\varphi_{\alpha,\beta}(z)=z\tilde\varphi_{\alpha,\beta}(z)$ and $B_\beta^\dagger\tilde\Psi_{\alpha,\beta}(z)=z\tilde\Psi_{\alpha,\beta}(z)$, as expected. However we find
$$
\frac{1}{\pi}\int_{\Bbb{C}}\,dz\, |\tilde\varphi_{\alpha,\beta}(z)><\tilde\Psi_{\alpha,\beta}(z)|=U(\alpha)\left(
\frac{1}{\pi}\int_{\Bbb{C}}\,dz\, |\varphi_0(z)><\varphi_0(z)|\,e^{z(\overline\alpha-\beta)+\overline z(\overline\beta-\alpha)}\right)
U(\overline\beta)^\dagger
$$
which returns $\1$ if $\alpha=\overline\beta$, but not otherwise. This is in agreement with Theorem \ref{thm1}: we have first seen that $\F_\varphi^{(\alpha,\beta)}$ and $\F_\Psi^{(\alpha,\beta)}$ are not Riesz bases. But they are biorthogonal. Hence the resolution of the identity for the associated coherent states needs not to be satisfied!

\section{The extended quantum harmonic oscillator}

The hamiltonian of this model, introduced in \cite{dapro}, is the  non self-adjoint operator $H_\beta=\frac{\beta}{2}\left(p^2+x^2\right)+i\sqrt{2}\,p$, where $\beta$ is a positive parameter and $[x,p]=i$. This hamiltonian is not ${\mc PT}$-symmetric but satisfies $\PP H_\beta=H_\beta^\dagger \PP$, where $\PP$ and $\mc T$ are the parity and the time-reversal operators. We will show that this hamiltonian produces 1-dimensional PB which are not regular.

Introducing the standard bosonic operators $a=\frac{1}{\sqrt{2}}\left(x+\frac{d}{dx}\right)$, $a^\dagger=\frac{1}{\sqrt{2}}\left(x-\frac{d}{dx}\right)$, $[a,a^\dagger]=\1$, and the number operator $N=a^\dagger a$, we can write $H_\beta=\beta N+(a-a^\dagger)+\frac{\beta}{2}\,\1$ which, defining now the operators
\be
\hat A_\beta=a-\frac{1}{\beta}, \qquad \hat B_\beta=a^\dagger+\frac{1}{\beta},
\label{31c}\en
can be written as
\be
H_\beta=\beta\left(\hat B_\beta \hat A_\beta+\gamma_\beta\,\1\right),
\label{32c}
\en
where $\gamma_\beta=\frac{2+\beta^2}{2\beta^2}$. It is clear that, for all $\beta>0$, $\hat A_\beta^\dagger\neq \hat B_\beta$ and that $[\hat A_\beta, \hat B_\beta]=\1$. Hence we have to do with pseudo-bosonic operators. This does not imply that Assumptions 1-4 hold true. But, comparing (\ref{32c}) with (\ref{sun1}), it is clear that $\hat A_\beta=A_{\frac{1}{\beta}}$ and $\hat B_\beta=B_{-\,\frac{1}{\beta}}$. So all the results of the previous section can be restated also for this model. In particular:
\begin{enumerate}

\item a non zero vector $\varphi_0^{(\beta)}\in\Hil$ exists such that $\hat A_\beta\varphi_0^{(\beta)}=0$ and $\varphi_0^{(\beta)}\in D^\infty(\hat B_\beta)$. This vector is a standard coherent state with parameter $\frac{1}{\beta}$:
$
\varphi_0^{(\beta)}=U(\beta^{-1})\varphi_0=e^{-1/2\beta^2}\,\sum_{k=0}^\infty\,\frac{\beta^{-k}}{\sqrt{k!}}\,\varphi_k,
$
where $\varphi_0$ is the vacuum of $a$, $a\varphi_0=0$, and $U(\beta^{-1})=e^{\frac{1}{\beta}(a^\dagger-a)}$ is the displacement operator already introduced before. Then $\varphi_n^{(\beta)}=\frac{1}{\sqrt{n!}}\,\hat B_\beta^n \varphi_0^{(\beta)}$ is a well defined vector for all $n\geq0$.

\item A non zero vector $\Psi_0^{(\beta)}\in\Hil$ exists such that $\hat B_\beta^\dagger\Psi_0^{(\beta)}=0$ and $\Psi_0^{(\beta)}\in D^\infty(\hat A_\beta^\dagger)$. This vector is also a  coherent state with parameter $-\frac{1}{\beta}$:
$
\Psi_0^{(\beta)}=\varphi_0^{(-\beta)}=U(-\beta^{-1})\varphi_0=U^{-1}(\beta^{-1})\varphi_0
$. Also, the vectors
$
\Psi_n^{(\beta)}=\frac{1}{\sqrt{n!}}\,(A_\beta^\dagger)^n \Psi_0^{(\beta)},
$
are well defined for all $n\geq 0$.

\item Let us now define the following sets of vectors: $\F_\varphi^{(\beta)}=\{\varphi_n^{(\beta)},\,n\geq 0\}$ and $\F_\Psi^{(\beta)}=\{\Psi_n^{(\beta)},\,n\geq 0\}$, their linear span $\D_\varphi^{(\beta)}$ and $\D_\Psi^{(\beta)}$, and the Hilbert spaces $\Hil_\varphi^{(\beta)}$ and $\Hil_\Psi^{(\beta)}$ obtained taking their closures. As in the previous section, we conclude that  $\Hil=\Hil_\Psi^{(\beta)}=\Hil_\varphi^{(\beta)}$.

\end{enumerate}

 General reasons discussed in \cite{bagpb1} show that, calling $\hat N_\beta=\hat B_\beta \hat A_\beta$ and $\hat \N_\beta=\hat N_\beta^\dagger=\hat A_\beta^\dagger \hat B_\beta^\dagger$, since
\be
\hat N_\beta\,\varphi_n^{(\beta)}=n\,\varphi_n^{(\beta)}, \qquad \hat \N_\beta \Psi_n^{(\beta)}= n\,\Psi_n^{(\beta)},
\label{36}\en
the vectors above are biorthogonal and, since $\left<\varphi_0^{(\beta)},\Psi_0^{(\beta)}\right>=e^{-2/\beta^2}$, the following holds:
\be
\left<\varphi_n^{(\beta)},\Psi_m^{(\beta)}\right>=\delta_{n,m} \,e^{-2/\beta^2}.
\label{37}\en

\vspace{2mm}

{\bf Remark:--} We could remove the factor $e^{-2/\beta^2}$ by changing the normalization of $\varphi_0^{(\beta)}$ and $\Psi_0^{(\beta)}$. We prefer to keep this normalization since it is standard for coherent states.

\vspace{2mm}

Of course, the results in the previous section also show that $\F_\varphi^{(\beta)}$ and $\F_\Psi^{(\beta)}$ are not Riesz bases for any choice of $\beta$. This same conclusion  can be deduced in a rather different way: we begin introducing the self-adjoint, unbounded and invertible operator: $V_\beta=e^{(a+a^\dagger)/\beta}$. Then, on the dense set $\E$,
\be
\hat A_\beta=V_\beta a V_\beta^{-1},\qquad \hat B_\beta=V_\beta a^\dagger V_\beta^{-1}.
\label{38}\en

Formula (\ref{38}) implies that $H_\beta$ can be related to a self adjoint operator $h_\beta=\beta(a^\dagger a+\gamma_\beta\1)$ as $H_\beta=V_\beta h_\beta V_\beta^{-1}$ or, equivalently as
\be
H_\beta V_\beta=V_\beta h_\beta,
\label{39}\en
which shows that $V_\beta$ is an intertwining operator (IO) relating $h_\beta$ and $H_\beta$. Moreover, taking the adjoint of (\ref{39}), we get $V_\beta H_\beta^\dagger=h_\beta V_\beta$, so that $V_\beta$ is also an IO between $h_\beta$ and $H_\beta^\dagger$. This has well known consequences on the spectra of the three operators $h_\beta$, $H_\beta$ and $H_\beta^\dagger$ and on their eigenstates, \cite{intop}.

In particular, recalling the expression of $\varphi_k$ in Section V, we have $h_\beta\varphi_k=\epsilon_k^{(\beta)}\varphi_k$, where $\epsilon_k^{(\beta)}=\beta(k+\gamma_\beta)$, $\forall k\geq0$. Hence, calling $\Phi_k^{(\beta)}:=V_\beta\,\varphi_k$,  using (\ref{39}) we have
$$
H_\beta \Phi_k^{(\beta)}= H_\beta V_\beta\,\varphi_k=V_\beta h_\beta \varphi_k=\epsilon_k^{(\beta)}V_\beta  \varphi_k=\epsilon_k^{(\beta)}\Phi_k^{(\beta)}.
$$
Moreover, because of (\ref{36}), since $H_\beta\,\varphi_k^{(\beta)}=\beta(\hat N_\beta+\gamma_\beta)\varphi_k^{(\beta)}=\epsilon_k^{(\beta)}\varphi_k^{(\beta)}$, and assuming that the eigenvalues $\epsilon_k^{(\beta)}$ are all non degenerate, it turns out that $\varphi_k^{(\beta)}=\alpha_k\,\Phi_k^{(\beta)}$ for all $k\geq0$, where $\alpha_k$ are simply complex constants. As a matter of fact we can further check that all these constants coincide: $\alpha_k=e^{-1/\beta^2}$, $k\geq0$,  so that, in conclusion,
\be
 \varphi_k^{(\beta)}=e^{-1/\beta^2}\,V_\beta \,\varphi_k,
 \label{310}\en
 for all $k\geq0$. Similar arguments, \cite{bagpb4}, show that
\be
 \Psi_k^{(\beta)}=e^{-1/\beta^2}\,V_\beta^{-1} \varphi_k,
 \label{310b}\en
for all $k\geq0$. This equation, together with (\ref{310}), also implies that $\Psi_k^{(\beta)}=V_\beta^{-2} \varphi_k^{(\beta)}$, $\forall k\geq0$. Therefore, recalling (\ref{213}), we recover the explicit expressions for the operators $S_\Psi^{(\beta)}$ and $S_\varphi^{(\beta)}$: $S_\Psi^{(\beta)}=V_\beta^{-2}$ and, consequently,  $S_\varphi^{(\beta)}=V_\beta^{2}$.

As a consequence, since the vectors  of the sets $\F_\varphi^{(\beta)}$ and $\F_\Psi^{(\beta)}$  are obtained by the o.n. basis $\E$ via the action of the two unbounded operators $V_\beta$ and $V_\beta^{-1}$, they are not Riesz bases. Hence, as expected, Assumption 4 in \cite{bagpb1} is not satisfied: we have PB which are not regular.

\section{The Swanson hamiltonian}

The starting point is the following non self-adjoint hamiltonian, \cite{dapro}:
$$
H_\theta=\frac{1}{2}\left(p^2+x^2\right)-\frac{i}{2}\,\tan(2\theta)\left(p^2-x^2\right),
$$
where $\theta$ is a real parameter taking value in $\left(-\frac{\pi}{4},\frac{\pi}{4}\right)\setminus\{0\}=:I$. It is clear that $H_\theta^\dagger=H_{-\theta}\neq H_\theta$, for all $\theta\in I$. As usual, $[x,p]=i\1$. Introducing the annihilation and creation operators $a$ and $a^\dagger$ we write
$$
H_\theta=N+\frac{i}{2}\,\tan(2\theta)\left(a^2+(a^\dagger)^2\right)+\frac{1}{2}\,\1,
$$
where $N=a^\dagger a$. This hamiltonian can be rewritten  by introducing the operators
\be
\left\{
\begin{array}{ll}
A_\theta=\cos(\theta)\,a+i\sin(\theta)\,a^\dagger,  \\
B_\theta=\cos(\theta)\,a^\dagger+i\sin(\theta)\,a,
\end{array}
\right.
\label{41c}\en
as
\be
H_\theta=\omega_\theta\left(B_\theta\,A_\theta+\frac{1}{2}\1\right),
\label{42}\en
where $\omega_\theta=\frac{1}{\cos(2\theta)}$ is well defined since $\cos(2\theta)\neq0$ for all $\theta\in I$. It is clear that $A_\theta^\dagger\neq B_\theta$ and that $[A_\theta,B_\theta]=\1$. To carry on now our analysis it is convenient to rewrite (\ref{41c}) by using the coordinate expressions for $a$ and $a^\dagger$:
\be
\left\{
\begin{array}{ll}
A_\theta=\frac{1}{\sqrt{2}}\left(e^{i\theta}x+e^{-i\theta}\,\frac{d}{dx}\right),  \\
B_\theta=\frac{1}{\sqrt{2}}\left(e^{i\theta}x-e^{-i\theta}\,\frac{d}{dx}\right).
\end{array}
\right.
\label{43y}\en
We are now ready to check the validity of Assumptions 1 and 2. We refer to \cite{bagpb4} for an {\em abstract} proof. Here we give the proof directly in coordinate.

Equation $A_\theta\varphi_0^{(\theta)}=0$ becomes $\left(e^{i\theta}x+e^{-i\theta}\frac{d}{dx}\right)\varphi_0^{(\theta)}(x)=0$ whose solution is
\be
\varphi_0^{(\theta)}(x)=N_1 \exp\left\{-\frac{1}{2}\,e^{2i\theta}\,x^2\right\},
\label{44y}\en
where $N_1$ is a normalization constant. Analogously, $B_\theta^\dagger\Psi_0^{(\theta)}=0$ becomes  $\left(e^{-i\theta}x+e^{i\theta}\frac{d}{dx}\right)\Psi_0^{(\theta)}(x)=0$, so that
\be
\Psi_0^{(\theta)}(x)=N_2 \exp\left\{-\frac{1}{2}\,e^{-2i\theta}\,x^2\right\},
\label{45y}\en
where, again, $N_2$ is a normalization constant. Notice that, since $\Re(e^{\pm 2i\theta})=\cos(2\theta)>0$ for all $\theta\in I$, both $\varphi_0^{(\theta)}(x)$ and $\Psi_0^{(\theta)}(x)$ belong to $\Lc^2({\Bbb R})$, which is the Hilbert space $\Hil$ of the theory.

Defining now the vectors $\varphi_n^{(\theta)}(x)$ and $\Psi_n^{(\theta)}(x)$ as in (\ref{22}), we find the following interesting result:

\be
\left\{
\begin{array}{ll}
\varphi_n^{(\theta)}(x)=\frac{1}{\sqrt{n!}}\,B_\theta^n\,\varphi_0^{(\theta)}(x)=\frac{N_1}{\sqrt{2^n\,n!}}
\,H_n\left(e^{i\theta}x\right)\,\exp\left\{-\frac{1}{2}\,e^{2i\theta}\,x^2\right\},  \\
\Psi_n^{(\theta)}(x)=\frac{1}{\sqrt{n!}}\,(A_\theta^\dagger)^n\,\Psi_0^{(\theta)}(x)=\frac{N_2}{\sqrt{2^n\,n!}}
\,H_n\left(e^{-i\theta}x\right)\,\exp\left\{-\frac{1}{2}\,e^{-2i\theta}\,x^2\right\},
\end{array}
\right.
\label{46}\en
where $H_n(x)$ is the n-th Hermite polynomial, \cite{bagpb4}.  The norm of these vectors can be given in terms of the Legendre polynomials $P_n$:
$$
\|\varphi_n^{(\theta)}\|^2=|N_1|^2\,\cos\left(\frac{\pi}{\cos(2\theta)}\right)\,
P_n\left(\frac{1}{\cos(2\theta)}\right)
$$
and
$$
\|\Psi_n^{(\theta)}\|^2=|N_2|^2\,\cos\left(\frac{\pi}{\cos(2\theta)}\right)\,
P_n\left(\frac{1}{\cos(2\theta)}\right),
$$
which are both well defined (even if the argument of $P_n$ does not belong to the interval $[-1,1]$), for all fixed $n$. Hence Assumptions 1 and 2 are satisfied.  What is not clear at this stage is whether the sets $\F_\varphi^{(\theta)}=\{\varphi_n^{(\theta)}(x),\,n\geq0\}$ and $\F_\Psi^{(\theta)}=\{\Psi_n^{(\theta)}(x),\,n\geq0\}$ are (i) complete in $\Lc^2(\Bbb{R})$; (ii) Riesz bases.

To answer to the first question we use the same general criterium adopted in Section IV, which was based on a result discussed in \cite{kolfom}.

We first notice that $\F_\varphi^{(\theta)}$ is complete in $\Lc^2(\Bbb{R})$ if and only if the set $\F_\pi^{(\theta)}:=\{\pi_n^{(\theta)}(x)=x^n\,
\exp\left\{-\frac{1}{2}\,e^{2i\theta}\,x^2\right\},\,n\geq0\}$ is complete in $\Lc^2(\Bbb{R})$. Hence, because of the above cited result and since $\exp\left\{-\frac{1}{2}\,e^{2i\theta}\,x^2\right\}$ satisfies for our values of $\theta$ the conditions required by the criterium, then $\F_\pi^{(\theta)}$ is complete and, as a consequence,
$\F_\varphi^{(\theta)}$ is complete in $\Lc^2(\Bbb{R})$. The same conclusion can be deduced for the set $\F_\Psi^{(\theta)}$, which is therefore also complete in $\Lc^2(\Bbb{R})$. Therefore, Assumption 3 is satisfied: $\Hil_\varphi=\Hil_\Psi=\Hil$.

Let us now go back to the biorthogonality of the two sets $\F_\varphi^{(\theta)}$ and $\F_\Psi^{(\theta)}$. Condition $\left<\varphi_0^{(\theta)},\Psi_0^{(\theta)}\right>=1$ is ensured by requiring that $\overline{N_1}\,N_2=\frac{e^{-i\theta}}{\sqrt{\pi}}$. Hence, with this choice, $\left<\varphi_n^{(\theta)},\Psi_m^{(\theta)}\right>=\delta_{n,m}$ which can be written explicitly as
$$
\int_{\Bbb{R}}H_n\left(e^{-i\theta}x\right)H_m\left(e^{-i\theta}x\right)e^{-e^{-2i\theta}x^2}\,dx=\delta_{n,m}\,\sqrt{2^{n+m}\,\pi\,n!\,m!}.
$$

To understand whether our biorthogonal sets are Riesz bases or not we  introduce the following unbounded, self-adjoint and invertible operator $T_\theta=e^{i\frac{\theta}{2}(a^2-{a^\dagger}^2)}$. Then we have
\be
A_\theta=T_\theta a T_\theta^{-1},\qquad B_\theta = T_\theta a^\dagger T_\theta^{-1}.
\label{47y}\en
This implies that $H_\theta=T_\theta h_\theta T_\theta^{-1}$, where $h_\theta=\omega_\theta\left(a^\dagger a+\frac{1}{2}\1\right)$. Hence, similarly to Section VI, we have deduced that $T_\theta$ is an IO:
\be
H_\theta T_\theta=T_\theta h_\theta,\qquad  T_\theta H_\theta^\dagger= h_\theta T_\theta.
\label{48c}\en
The same arguments discussed in the previous section show that, if the eigenvalues $\omega_n^{(\theta)}=\omega(n+1/2)$ are non degenerate, then a single complex constant $\alpha$ must exist such that
\be
\varphi_n^{(\theta)}=\alpha\, T_\theta\, \varphi_n, \quad \mbox{ and }\quad \Psi_n^{(\theta)}=\frac{1}{\overline\alpha}\, T_\theta^{-1}\, \varphi_n,
\label{49}\en
$\varphi_n\in\E$. These equalities show, in particular, that neither $\F_\varphi^{(\theta)}$ nor $\F_\Psi^{(\theta)}$ are Riesz bases. Also, we deduce that $S_\varphi^{(\theta)}=|\alpha|^2\,T_\theta^2$ and $S_\Psi^{(\theta)}=|\alpha|^{-2}\,T_\theta^{-2}$.

\vspace{2mm}

A two-parameters extension of this model has also been introduced in \cite{bagpb4}, and an even more general extension is discussed in \cite{baglast}.

\section{Generalized Landau levels (GLL)}

The Hamiltonian of a single electron, moving on a two-dimensional
plane and subject to a uniform magnetic field along the positive
$z$-direction, is given by the operator
\begin{equation}
H_0'={\frac 12}\,\left(\underline p+\underline A(r)\right)^2={\frac
12}\; \left(p_x-{\frac y2}\right)^2+{\frac 12}\,\left(p_y+{\frac
x2} \right)^2, \label{41}
\end{equation}
where we have used minimal coupling and the symmetric gauge $\vec
A=\frac{1}{2}(-y,x,0)$.

The spectrum of this Hamiltonian is easily obtained by first
introducing the new variables
  \be
\label{42d}
  P_0'= p_x-y/2, \hspace{5mm}     Q_0'= p_y+x/2.
  \en
In terms of $P_0'$ and $Q_0'$ the  hamiltonian
$H_0$ can be rewritten as
 \be
\label{43d}
  H_0'=\frac{1}{2}(Q_0'^2 + P_0'^2).
  \en
On a classical level, the transformation (\ref{42d}) is part of a canonical map from the
phase space variables $(x,y,p_x,p_y)$ to $(Q_0,P_0,Q_0',P_0')$,
where
 \be
\label{44d}
   P_0= p_y-x/2, \hspace{5mm}
  Q_0= p_x+y/2,
   \en
which can be used to construct a second hamiltonian $ H_0=\frac{1}{2}(Q_0^2 + P_0^2)$, which describes  an electron  moving on a two-dimensional
plane and subject to a uniform magnetic field along the negative
$z$-direction.

  The corresponding quantized  operators  satisfy the commutation relations:
$$ [x, p_x] = [y, p_y] = i, \quad [x,p_y] = [y,p_x] = [x,y] = [p_x , p_y ] = 0, $$
and
  \be
\label{45d}
 [Q_0,P_0] = [Q_0',P_0']=i, \quad  [Q_0,P_0']=[Q_0',P_0]=[Q_0,Q_0']=[P_0,P_0']=0,
  \en
so that $[H_0,H_0']=0$.

In \cite{alibag} we have considered, in the context of supersymmetric
(SUSY) quantum mechanics, a natural extension of $H_0'$:  introducing the vector valued function $\vec
W_0=-\frac{1}{2}(x,y,0)=(W_{0,1},W_{0,2},0)$, we may rewrite the
operators in (\ref{42d}) and (\ref{44d}) as \be
P_0'=p_x+W_{0,2},\hspace{4mm}Q_0'=p_y-W_{0,1},\hspace{4mm}
P_0=p_y+W_{0,1},\hspace{4mm}Q_0=p_x-W_{0,2}.\label{48d}\en These
definitions were extended in \cite{alibag} as follows: \be
p'=p_x+W_{2},\hspace{4mm}q'=p_y-W_{1},\hspace{4mm}
p=p_y+W_{1},\hspace{4mm}q=p_x-W_{2},\label{49d}\en introducing a
vector superpotential $\vec W=(W_{1},W_{2},0)$.
Here, since we are interested in constructing 2-d pseudo-bosons, we introduce two (in general) complex and different vector
superpotentials (this is a slight abuse of language!) $\vec
W=(W_1,W_2)$ and $\vec V=(V_1,V_2)$, and we put \be
P'=p_x+W_{2},\hspace{4mm}Q'=p_y-W_{1},\hspace{4mm}
P=p_y+V_{1},\hspace{4mm}Q=p_x-V_{2}.\label{410d}\en Our notation is
the following: all operators with suffix $0$ are related to
the standard Landau levels (SLL). The same operators, without the $0$, refer to our
generalized model, i.e. to the GLL. Notice that these operators are, in general, not self-adjoint. Hence, while for instance $P_0=P_0^\dagger$, we may have $P\neq P^\dagger$, depending on the choice of $V_1$.  The superpotentials should
 also be chosen in such a way that, first of all, $Q$, $P$, $Q'$ and
$P'$ satisfy the following commutation rules:
  \be
\label{412d}
 [Q,P] = [Q',P']=i, \quad  [Q,P']=[Q',P]=[Q,Q']=[P,P']=0.
  \en
These impose certain conditions on $\vec V$ and $\vec W$: \be
W_{1,x}=V_{2,y},\quad W_{2,x}=-V_{2,x},\quad
W_{1,y}=-V_{1,y},\quad W_{2,y}=V_{1,x}, \label{413d}\en
as well as
\be V_{1,x}+V_{2,y}=W_{1,x}+W_{2,y}=-1. \label{414d}\en
The subscripts $x,y$ denote differentiation with
respect to that variable.
Hence, as
it was already clear at the beginning, the two different vector
superpotentials must be related to each other.  We now
introduce \be A'=\alpha'(Q'+i\,P'),\quad
B'=\gamma'(Q'-i\,P'),\quad A=\alpha (Q+i\, P),\quad B=\gamma (Q -i
P), \label{415d}\en where $\alpha\,\gamma=\frac{1}{2}$ and
$\alpha'\,\gamma'=\frac{1}{2}$.  Thus, the operators generalizing the Landau
Hamiltonians in \cite{alibag} are \be h'={\frac 12}\;
\left(p_x+W_2\right)^2+{\frac 12}\,\left(p_y-W_1\right)^2, \qquad
h={\frac 12}\; \left(p_x-V_2\right)^2+{\frac
12}\,\left(p_y+V_1\right)^2, \label{415db}\en which can be
rewritten as \be h'=B'A'-\frac{1}{2}\,\1,\qquad
h=BA-\frac{1}{2}\1. \label{416d}\en The operators in (\ref{415d})
are pseudo-bosonic since they satisfy the following commutation
rules: \be [A,B]=[A',B']=\1,\label{417d}\en while all the other
commutators are zero. It is important to observe that, in general, $B\neq
A^\dagger$ and $B'\neq {A'}^\dagger$.

The following are some possible choices of $\vec W$ and $\vec V$ satisfying (\ref{413d}) and (\ref{414d}):

\vspace{2mm}

{\bf Choice 1} (SLL). Let us take
$V_1(x,y)=W_1(x,y)=-\frac{x}{2}$, \quad
$V_2(x,y)=W_2(x,y)=-\frac{y}{2}$. If we further take
$\alpha=\gamma=\alpha'=\gamma'=\frac{1}{\sqrt{2}}$ we recover
exactly the usual situation, \cite{alibag}. Moreover, we go back to bosonic rather than pseudo-bosonic commutation relations.

\vspace{2mm}

{\bf Choice 2} (Perturbations of the SLL). First we consider a
symmetric perturbation. For that we take
$V_1(x,y)=-\frac{x}{2}+v_1(y)$, $V_2(x,y)=-\frac{y}{2}+v_2(x)$,
where $v_1$ and $v_2$ are arbitrary (but sufficiently regular)
functions. Hence we get, apart from unessential additive constants,
$W_1(x,y)=-\frac{x}{2}-v_1(y)$, $W_2(x,y)=-\frac{y}{2}-v_2(x)$.
In order not to  trivialize the situation, it is also necessary to
take $v_1(y)$ and $v_2(x)$ complex (at least one of them): this is
the way to get PB rather than {\em simple} bosons.

An asymmetric (in $x$ and $y$) version of this perturbation can be constructed by
just taking $V_1(x,y)=-a_1\,x+v_1(y)$, $V_2(x,y)=-a_2\,y+v_2(x)$,
with $a_1+a_2=1$.

\vspace{2mm}

{\bf Choice 3} (A general solution). We take
$V_1(x,y)=-x+v_1(y)+\int \frac{\partial V_2(x,y)}{\partial
y}\,dx$, where $V_2(x,y)$ is any function for which this
definition makes sense. In particular, for instance, if we take
$V_2(x,y)=e^{xy}$ then
$V_1(x,y)=-x+v_1(y)+\frac{1}{y^2}\left(x\,y-1\right)e^{xy}$ and,
consequently,
$W_1(x,y)=-v_1(y)-\frac{1}{y^2}\left(x\,y-1\right)e^{xy}$ and
$W_2(x,y)=-y-e^{xy}$.

If we rather take $V_2(x,y)=x^n\,y^k$, $n,k=1,2,3,\ldots$, then
$V_1(x,y)=-x+v_1(y)-\frac{k}{n+1}\,x^{n+1}y^{k-1}$, and so on.

\subsection{A perturbation of the SLL}

We will now focus our attention on Choice 2 above, with an
explicit choice of $v_1(y)$ and $v_2(x)$, and apply the
construction given in Section II. Let
\be
W_1(x,y)=-\frac{x}{2}-ik_1y,\qquad
W_2(x,y)=-\frac{y}{2}-ik_2x,\label{418}
\en
with $k_1$ and $k_2$
real and not both zero (not to go back to
 SLL). In this case the operators in (\ref{415d}) assume the following differential expressions: \bea
 \left\{
    \begin{array}{ll}
A'=\alpha'\left(\partial_x-i\partial_y+\frac{x}{2}(1+2k_2)-\frac{iy}{2}(1-2k_1)\right),\\
B'=\gamma'\left(-\partial_x-i\partial_y+\frac{x}{2}(1-2k_2)+\frac{iy}{2}(1+2k_1)\right),\\
A=\alpha\left(-i\partial_x+\partial_y-\frac{ix}{2}(1+2k_2)+\frac{y}{2}(1-2k_1)\right),\\
B=\gamma\left(-i\partial_x-\partial_y+\frac{ix}{2}(1-2k_2)+\frac{y}{2}(1+2k_1)\right).\\
      \end{array}
        \right.\label{419} \ena
In order to check Assumptions 1 and 2 of the previous section, we
first look for vectors $\varphi_{0,0}(x,y)$ and $\Psi_{0,0}(x,y)$
satisfying $A\varphi_{0,0}(x,y)=0$ and
$B^\dagger\Psi_{0,0}(x,y)=0$. We get \bea
 \left\{
    \begin{array}{ll}
\varphi_{0,0}(x,y)=N_\varphi\,\exp\left\{-\frac{x^2}{4}(1+2k_2)-\frac{y^2}{4}(1-2k_1)\right\}\\
\Psi_{0,0}(x,y)=N_\Psi\,\exp\left\{-\frac{x^2}{4}(1-2k_2)-\frac{y^2}{4}(1+2k_1)\right\},\\
\end{array}\right.
\label{420}\ena where $N_\varphi$ and $N_\Psi$ are normalization
constants which we fix in such a way that
$\left<\varphi_{0,0},\Psi_{0,0}\right>=1$. Of course, in order for
this result to make sense, the two functions must belong to the
Hilbert space $\Hil$ we are considering here, i.e.
$\Lc^2(\Bbb{R}^2)$. This imposes some constraints on $k_1$ and
$k_2$: $-\frac{1}{2}<k_j<\frac{1}{2}$, $j=1,2$.

It is possible to check that the same functions also satisfy
$A'\varphi_{0,0}(x,y)=0$ and $B'^\dagger\Psi_{0,0}(x,y)=0$. It is
now evident that Assumptions 1 and 2 are satisfied. Indeed the
action of, say, $B_1^n$  on $\varphi_{0,0}(x,y)$ simply produces
some polynomial, \cite{abg}, of the $n$-th degree times
a gaussian: this resulting function belongs clearly to
$\Lc^2(\Bbb{R}^2)$ for all $n$. Then we can introduce the following
functions \be
\varphi_{n,l}(x,y)=\frac{B'^n\,B^l}{\sqrt{n!\,l!}}\,\varphi_{0,0}(x,y),
\quad\mbox{ and }\quad
\Psi_{n,l}(x,y)=\frac{(A'^\dagger)^n\,(A^\dagger)^l}{\sqrt{n!\,l!}}\,\Psi_{0,0}(x,y),
\label{421}\en where $n,l=0,1,2,3,\ldots$. As we have seen in the
previous section, the sets
$\F_\Psi=\{\Psi_{n,l}(x,y),\,n,l\geq0\}$ and
$\F_\varphi=\{\varphi_{n,l}(x,y),\,n,l\geq0\}$ are biorthogonal.
In fact, with our previous choice of the normalization constants,
we have \be
\left<\Psi_{n,l},\varphi_{m,k}\right>=\delta_{n,m}\delta_{l,k},
\quad \forall n,m,l,k\geq0. \label{422}\en Of course these vectors
diagonalize the operators $h=N-\frac{1}{2}\,\1$ and
$h'=N'-\frac{1}{2}\,\1$, as well as their adjoints
$h^\dagger=\N-\frac{1}{2}\,\1$ and
$h'^\dagger=\N'-\frac{1}{2}\,\1$, where $N=BA$, $N'=B'A'$,
$\N=N^\dagger$ and $\N'=N'^\dagger$. We find:
$$
h'\varphi_{n,l}=\left(n-\frac{1}{2}\right)\varphi_{n,l}, \quad
h\,\varphi_{n,l}=\left(l-\frac{1}{2}\right)\varphi_{n,l},
$$
and
$$
h'^\dagger\Psi_{n,l}=\left(n-\frac{1}{2}\right)\Psi_{n,l}, \quad
h^\dagger\Psi_{n,l}=\left(l-\frac{1}{2}\right)\Psi_{n,l}.
$$
The two sets $\F_\varphi$ and
$\F_\Psi$ are complete in $\Hil$: this is a consequence of the
fact that (a.) the set
$\F_h:=\{h_{n,m}(x,y):=x^n\,y^m\,\varphi_{0,0}(x,y), \,n,m\geq0\}$
is complete in $\Lc^2(\Bbb{R}^2)$; (b.) each function of $\F_h$,
can be written as a finite linear combination of some
$\varphi_{i,j}(x,y)$.

This  result implies that also Assumption 3 of Section II is
satisfied. Now we could introduce the intertwining operators
$S_\varphi$ and $S_\Psi$ and check, among other properties, if
they are bounded or not. We  introduce first the o.n. basis of
$\Lc^2(\Bbb{R}^2)$ arising from the SLL, \cite{alibag},
$$\F_{\varphi}^{(0)}:=\left\{\varphi_{n,l}^{(0)}(x,y):=\frac{B_0'^n\,B_0^l}{\sqrt{n!\,l!}}
\varphi_{0,0}^{(0)}(x,y),\quad n,m\geq0\right\},$$ where
$\varphi_{0,0}^{(0)}(x,y)=\frac{1}{\sqrt{2\pi}}\,e^{-(x^2+y^2)/4}$
is the vacuum of $A_0=\frac{1}{\sqrt{2}}(Q_0+iP_0)$ and
$A_0'=\frac{1}{\sqrt{2}}(Q_0'+iP_0')$. Recall that, for SLL,
$B'_0=A_0'^\dagger$ and $B_0=A_0^\dagger$.

To prove now that $\F_\varphi$ is not a Riesz basis, it is enough to show
that an invertible operator $T_\varphi$ exists, mapping $\F_{\varphi}^{(0)}$
into $\F_{\varphi}$,   but
$T_\varphi$ and/or $T_\varphi^{-1}$ are not bounded. In \cite{abg} we have deduced the analytic expression for $T_\varphi$, which is just the following multiplication operator:
\be
T_\varphi=\frac{\varphi_{0,0}(x,y)}{\varphi_{0,0}^{(0)}(x,y)}=\sqrt{2\pi}N_\varphi\,
e^{-\frac{x^2}{2}\,k_2+\frac{y^2}{2}\,k_1}. \label{426}\en The inverse of $T_\varphi$ is
$T_\varphi^{-1}=\frac{1}{\sqrt{2\pi}N_\varphi}\,e^{\frac{x^2}{2}\,k_2-\frac{y^2}{2}\,k_1}$.
It is clear that both $T_\varphi$ and/or $T_\varphi^{-1}$ are
unbounded on $\Lc^2(\Bbb{R}^2)$ for all possible choices of $k_1$ and $k_2$ in $\left(-\,\frac{1}{2},\frac{1}{2}\right)$, except when $k_1 = k_2 = 0$, i.e., in the case of the SLL. Hence, for well known general
reasons, \cite{you,chri}, $\F_\varphi$ cannot be a Riesz basis.

Essentially the same arguments  also show that $\F_\Psi$ is not a
Riesz basis, either. Indeed, an operator $T_\Psi$ mapping
$\F_{\varphi}^{(0)}$ into $\F_{\Psi}$ can be found and it is \be
T_\Psi=\frac{\Psi_{0,0}(x,y)}{\varphi_{0,0}^{(0)}(x,y)}=\sqrt{2\pi}N_\Psi\,e^{\frac{x^2}{2}\,k_2-\frac{y^2}{2}\,k_1}.
\label{428}\en This operator satisfies
$\Psi_{n,l}(x,y)=T_\Psi\varphi_{n,l}^{(0)}(x,y)$ for all possible
choices of $n$ and $l$ greater or equal to zero. Therefore, since
$\varphi_{n,l}(x,y)=T_\varphi\varphi_{n,l}^{(0)}(x,y)= (T_\varphi
T_\Psi^{-1})\Psi_{n,l}(x,y)$, the operators $S_\varphi$ and
$S_\Psi$ in (\ref{213}) can be easily identified and look like \be
S_\varphi=T_\varphi
T_\Psi^{-1}=\frac{N_\varphi}{N_\Psi}\,e^{-x^2k_2+y^2k_1}, \quad
S_\Psi=S_\varphi^{-1}=T_\Psi
T_\varphi^{-1}=\frac{N_\Psi}{N_\varphi}\,e^{x^2k_2-y^2k_1}.
\label{430}\en Notice that for any choice of $k_1$ and $k_2$  in
$\left(-\frac{1}{2},\frac{1}{2}\right)$, other than  $(k_1 , k_2 ) = (0,0)$, at least one of these
operators is unbounded.

\vspace{3mm}

We refer to \cite{abg} for a brief analysis of bi-coherent states related to this model.

\vspace{3mm}

An alternative way to construct examples of pseudo-bosons out of the Landau levels has been discussed in \cite{bagbenasque}. The idea is quite simple: rather than replacing the operators $(Q_0,P_0,Q_0',P_0')$ with $(Q,P,Q',P')$, as we have done before, we simply {\em translate} (some of) the operators $X_0$, mimicking what we did in Section V. Moreover, in the same paper, we have also proven the following no-go result: suppose $a$ and $a^\dagger$ are two operators acting on $\Hil$ and satisfying $[a,a^\dagger]=\1$. Then, for all $\alpha\neq0$, the operators $A:=a-\alpha {a^\dagger}^2$ and $B:=a^\dagger$ are such that $[A,B]=\1$, $A^\dagger\neq B$, but they do not satisfy Assumption 1.

In fact, if such a non zero vector $\varphi_0\in\Hil$ exists, then it could be expanded in terms of the eigenvectors $\Phi_n:=\frac{{a^\dagger}^n}{\sqrt{n!}}\Phi_0$, $a\Phi_0=0$, of the number operator $N=a^\dagger a$: $\varphi_0=\sum_{n=0}^\infty c_n\Phi_n$, for some sequence $\{c_n, n\geq0\}$ such that $\sum_{n=0}^\infty|c_n|^2<\infty$. Condition $A\varphi_0=0$ can be rewritten as $a\varphi_0=\alpha {a^\dagger}^2\varphi_0$. Now, inserting in both sides of this equality the expansion for $\varphi_0$, and recalling  that $a^\dagger\Phi_n=\sqrt{n+1}\Phi_{n+1}$ and $a\Phi_n=\sqrt{n}\Phi_{n-1}$, $n\geq0$, we deduce the following relations between the coefficients $c_n$: $c_1=c_2=0$ and $c_{n+1}\sqrt{n+1}=\alpha c_{n-2}\sqrt{(n-1)n}$, for all $n\geq2$. The solution of this recurrence relation is the following:
$$
c_3=\alpha c_0\frac{\sqrt{3!}}{3},\quad c_6=\alpha^2 c_0\frac{\sqrt{6!}}{3\cdot6}, \quad c_9=\alpha^3 c_0\frac{\sqrt{9!}}{3\cdot6\cdot9}, \quad c_{12}=\alpha^4 c_0\frac{\sqrt{12!}}{3\cdot6\cdot9\cdot12},
$$
and so on. Then
$$
\varphi_0=c_0\left(\Phi_0+\sum_{k=1}^\infty\,\alpha^k\frac{\sqrt{(3k)!}}{1\cdot3\cdots3k}\,\Phi_{3k}\right)
$$
However, computing $\|\varphi_0\|$ we deduce that this series only converge if $\alpha=0$, i.e. if $A$ coincides with $a$ and $B$ with $a^\dagger$.

\vspace{2mm}

A similar results can be obtained considering the operators $A:=a-\alpha {a^\dagger}^n$ and $B:=a^\dagger-\beta\1$, $n\geq2$, $\alpha, \beta\in \Bbb{C}$. Again we find $[A,B]=\1$, $A^\dagger\neq B$, and again, with similar techniques, we deduce that they do not satisfy Assumption 1. In the same way, if we define $A:=a-\alpha \1$ and $B:=a^\dagger-\beta a^m$, $m\geq2$, $\alpha, \beta\in \Bbb{C}$, we find that, in general, $[A,B]=\1$, $A^\dagger\neq B$, but they do not satisfy Assumption 2. This  suggests that  if we try to define, starting from $a$ and $a^\dagger$, new operators $A=a+f(a,a^\dagger)$ and $B=a^\dagger+g(a,a^\dagger)$, only very special choices of $f$ and $g$ are compatible with the pseudo-bosonic structure outlined in Section II.

\section{Damped harmonic oscillator}

An interesting example of formal two-dimensional PB is
provided by the damped harmonic oscillator (DHO), see \cite{ban} or \cite{chru} for a quantum approach to the system.
To use a lagrangian approach for the DHO the original
equation of motion, $m\ddot x+\gamma \dot x+kx=0$, is
complemented by a second {\em virtual} equation,  $m\ddot y-\gamma
\dot y+ky=0$, and the classical lagrangian for the system looks
like $L=m\dot x\dot y+\frac{\gamma}{2}(x\dot y-\dot xy)-kxy$,
which corresponds to a classical Hamiltonian $H=p_x\,\dot
x+p_y\,\dot
y-L=\frac{1}{m}\left(p_x+\gamma\frac{y}{2}\right)\left(p_y-\gamma\frac{x}{2}\right)+kxy$,
where $p_x=\frac{\partial L}{\partial\dot x}$ and
$p_y=\frac{\partial L}{\partial\dot y}$ are the conjugate momenta.
The introduction of pseudo-bosons is based on two successive
changes of variables and on a canonical quantization. First of all
we introduce the new variables $x_1$ and $x_2$ via
$x=\frac{1}{\sqrt{2}}(x_1+x_2)$, $y=\frac{1}{\sqrt{2}}(x_1-x_2)$.
Then $L=\frac{1}{2}m\left(\dot x_1^2-\dot
x_2^2\right)+\frac{\gamma}{2}\left(x_2\dot x_1-x_1\dot
x_2\right)-\frac{k}{2}(x_1^2-x_2^2)$ and
$H=\frac{1}{2m}\left(p_1-\gamma\frac{x_2}{2}\right)^2+\frac{1}{2m}\left(p_2+\gamma\frac{x_1}{2}\right)^2+
\frac{k}{2}(x_1^2-x_2^2)$. The second change of variable is the
following:

\bea
 \left\{
    \begin{array}{ll}
p_+=\sqrt{\frac{\omega_+}{2m\Omega}}p_1+i\,\sqrt{\frac{m\Omega\omega_+}{2}}\,x_2,\\
p_-=\sqrt{\frac{\omega_-}{2m\Omega}}p_1-i\,\sqrt{\frac{m\Omega\omega_-}{2}}\,x_2,\\
x_+=\sqrt{\frac{m\Omega}{2\omega_+}}x_1+i\,\sqrt{\frac{1}{2m\Omega\omega_+}}\,p_2,\\
x_-=\sqrt{\frac{m\Omega}{2\omega_-}}x_1-i\,\sqrt{\frac{1}{2m\Omega\omega_-}}\,p_2,\\
      \end{array}
        \right.\label{51} \ena
where we have introduced
$\Omega=\sqrt{\frac{1}{m}\left(k-\frac{\gamma^2}{4m}\right)}$ and
the  two following complex quantities $\omega_\pm=\Omega\pm
i\frac{\gamma}{2m}$. For simplicity we will assume
that $k\geq \frac{\gamma^2}{4m}$, so that $\Omega$ is
real. Up to now, we are still at a classical level, so that
$\overline\omega_+=\omega_-$, $\overline p_+=p_-$, $\overline
x_+=x_-$, and consequently, see below, $\overline H_+=H_-$ and $\overline H=H$. Hence
$H$ is a real Hamiltonian. Indeed, with these definitions, the Hamiltonian
looks like the hamiltonian of a two-dimensional harmonic
oscillator
$$
H=\frac{1}{2}\left(p_+^2+\omega_+^2x_+^2\right)+\frac{1}{2}\left(p_-^2+\omega_-^2x_-^2\right)=:H_++H_-
$$
at least formally.

At this stage we quantize canonically  the system,
\cite{ban}, requiring that the following commutators are
satisfied: \be [x_+,p_+]=[x_-,p_-]=i, \label{52}\en all the other
commutators being trivial. We also have to require that $
p_+^\dagger=p_-$ and that $x_+^\dagger=x_-$, which are the quantum
version of the {\em compatibility} conditions above.  The
pseudo-bosons now appear:
 \bea
 \left\{
    \begin{array}{ll}
a_+=\sqrt{\frac{\omega_+}{2}}\left(x_++i\,\frac{p_+}{\omega_+}\right),\\
a_-=\sqrt{\frac{\omega_-}{2}}\left(x_-+i\,\frac{p_-}{\omega_-}\right),\\
b_+=\sqrt{\frac{\omega_+}{2}}\left(x_+-i\,\frac{p_+}{\omega_+}\right),\\
b_-=\sqrt{\frac{\omega_-}{2}}\left(x_--i\,\frac{p_-}{\omega_-}\right),\\
      \end{array}
        \right.\label{53} \ena
and indeed we have $[a_+,b_+]=[a_-,b_-]=\1$, all the other
commutators being zero. Notice also that $b_+=a_-^\dagger$ and
$b_-=a_+^\dagger$. Moreover $H$ can be written in term of the
operators $N_\pm=b_\pm a_\pm$ as
$H=\omega_+N_++\omega_-N_-+\frac{\omega_++\omega_-}{2}\,\1$. Hence the
hamiltonian of the quantum DHO is easily
written in terms of pseudo-bosonic operators.

\vspace{2mm}

This system provides a non trivial example of pseudo-bosonic
operators which do not satisfy any of the Assumptions 1-3 of
Section II. To show this, we first observe that a possible
representation of the operators in (\ref{52}) satisfying the compatibility conditions $ p_+^\dagger=p_-$ and
$x_+^\dagger=x_-$ is the following \bea
 \left\{
    \begin{array}{ll}
x_+=\frac{1}{\Gamma\,\overline{\delta}-\delta\,\overline{\Gamma}}\left(\overline{\Gamma}\,p_y+\overline{\delta}\,x\right),\\
x_-=\frac{-1}{\Gamma\,\overline{\delta}-\delta\,\overline{\Gamma}}\left({\Gamma}\,p_y+{\delta}\,x\right),\\
p_+=\Gamma \,p_x+\delta \,y,\\
p_-=\overline{\Gamma} \,p_x+\overline{\delta} \,y,\\
      \end{array}
        \right.\label{54} \ena
for all choices of $\Gamma$ and $\delta$ such that
$\Gamma\,\overline{\delta}\neq \delta\,\overline{\Gamma}$. Here
$x$, $y$, $p_x$ and $p_y$ are pairwise conjugate self-adjoint
operators: $[x,p_x]=[y,p_y]=\1$.  Hence, representing $x$ and $y$ as
the standard multiplication operators and $p_x$ and $p_y$ as
$-i\,\frac{\partial}{\partial\,x}=-i\,\partial_x$ and
$-i\,\frac{\partial}{\partial\,y}=-i\,\partial_y$,  we get

\bea
 \left\{
    \begin{array}{ll}
a_+=\sqrt{\frac{\omega_+}{2}}\,\left\{\left(\beta\,x+i\,\frac{\delta}{\omega_+}\,y\right)+\left(\frac{\Gamma}{\omega_+}\,
\partial_x-i\,\alpha\,\partial_y\right)\right\},\\
a_-=\sqrt{\frac{\omega_-}{2}}\,\left\{\left(\overline{\beta}\,x+i\,\frac{\overline{\delta}}{\omega_-}\,y\right)+
\left(\frac{\overline{\Gamma}}{\omega_-}\,\partial_x-i\,\overline{\alpha}\,\partial_y\right)\right\},\\
b_+=\sqrt{\frac{\omega_+}{2}}\,\left\{\left(\beta\,x-i\,\frac{\delta}{\omega_+}\,y\right)-\left(\frac{\Gamma}{\omega_+}\,
\partial_x+i\,\alpha\,\partial_y\right)\right\},\\
b_-=\sqrt{\frac{\omega_-}{2}}\,\left\{\left(\overline{\beta}\,x-i\,\frac{\overline{\delta}}{\omega_-}\,y\right)-
\left(\frac{\overline{\Gamma}}{\omega_-}\,\partial_x+i\,\overline{\alpha}\,\partial_y\right)\right\},\\
      \end{array}
        \right.\label{55} \ena
where, to simplify the notation, we have introduced
$\alpha=\frac{\overline{\Gamma}}{\Gamma\,\overline{\delta}-\delta\,\overline{\Gamma}}$
and
$\beta=\frac{\overline{\delta}}{\Gamma\,\overline{\delta}-\delta\,\overline{\Gamma}}$.

Assumption 1 of Section II requires in particular the existence of a
square-integrable function $\varphi_{0,0}(x,y)$ such that, first
of all, $a_+\varphi_{0,0}(x,y)=a_-\varphi_{0,0}(x,y)=0$.
But such  a solution  is easily found:
\be
\varphi_{0,0}(x,y)=N_0\,\exp\left\{-\,\frac{\beta\,\omega_+}{2\,\Gamma}\,x^2+\frac{\delta}{2\,\alpha\,\omega_+}\,y^2\right\},
\label{56}\en where
$\frac{\omega_+}{\omega_-}=-\,\frac{\delta}{\overline{\delta}}\,\frac{\Gamma}{\overline{\Gamma}}$.
Now, in order for $\varphi_{0,0}(x,y)$ to be square integrable, we need to require
that
$\Re\left(\frac{\beta\,\omega_+}{2\Gamma}\right)=\frac{\beta\,\omega_+}{2\Gamma}>0$ and $\Re\left(\frac{\delta}{\alpha\,\omega_+}\right)=\frac{\delta}{\alpha\,\omega_+}<0$.
Now, it is not hard to check that these two conditions are
incompatible: if one is verified, the other is not. Therefore the
conclusion is that, following the procedure we have considered so
far, Assumptions 1 is violated. Analogously, we could check that Assumption 2 is violated as well and, of course, Assumption 3
cannot even be considered since it is meaningless. Hence for our pseudo-bosonic operators the procedure discussed in Section II does not work at all. This is a non trivial example showing that the commutation rule in (\ref{21}) is not enough to  produce any {\em interesting} functional settings.

\subsection{Changing Hilbert space}

This result does not exclude, a priori, that a solution of Assumptions 1 and 2 could be found in a different $\Lc^2$ space, for instance in a space with
a suitable weight. This possibility was considered in \cite{DHO}, where $\Lc^2({\Bbb R}^2)$ as been replaced by $\Hil_1:=\Lc^2\left({\Bbb
R}^2,e^{-c_1x^2-c_2y^2}\,dx\,dy\right)$, where $c_1$ and $c_2$ are two positive constants chosen in such a way that the wave-function
$\varphi_{0,0}(x,y)$ in (\ref{56}) does belong to $\Hil_1$. However, in $\Hil_1$ the adjoint of the
operator is different from the one in $\Lc^2({\Bbb R}^2)$, and this should be taken into account to produce a consistent model.

Indeed, in $\Hil_1$ the scalar product is clearly defined as follows: $$\left<f,g\right>_1=\int_{\Bbb R}dx\int_{\Bbb
R}dy\,\overline{f(x)}\,g(x)\,e^{-c_1x^2-c_2y^2}.$$ The adjoint $X^*$ of the operator $X$ in $\Hil_1$,  is defined by the equation
$\left<Xf,g\right>_1=\left<f,X^*g\right>_1$, for all $f, g\in\Hil_1$ belonging respectively to the domain of $X$ and $X^*$.  It is easy to check that
$\partial_x^*=-\partial_x+2c_1x$ and $\partial_y^*=-\partial_y+2c_2y$. Taking
as our starting point the operators $a_\pm$ and $b_\pm$ defined in (\ref{55}), we can compute their {\em new} adjoints (i.e. their adjoints in
$\Hil_1$), which can be written as \bea
 \left\{
    \begin{array}{ll}
a_+^*=b_-+\sqrt{2\omega_-}\left(c_1x\,\frac{\overline\Gamma}{\omega_-}+ic_2y\overline{\alpha}\right),\\
a_-^*=b_++\sqrt{2\omega_+}\left(c_1x\,\frac{\Gamma}{\omega_+}+ic_2y{\alpha}\right),\\
b_+^*=a_-+\sqrt{2\omega_-}\left(-c_1x\,\frac{\overline\Gamma}{\omega_-}+ic_2y\overline{\alpha}\right),\\
b_-^*=a_++\sqrt{2\omega_+}\left(-c_1x\,\frac{\overline\Gamma}{\omega_-}+ic_2y{\alpha}\right).\\
      \end{array}
        \right.\label{71} \ena
It is clear that, but for this case, the
compatibility conditions required above are not satisfied: $a_\pm^*\neq b_\mp$. Nevertheless, if we carry on our analysis, we can still
look for the solutions of the differential equations $a_+\varphi_{0,0}(x,y)=a_-\varphi_{0,0}(x,y)=0$ and
$b_+^*\Psi_{0,0}(x,y)=b_-^*\Psi_{0,0}(x,y)=0$. The solution $\varphi_{0,0}(x,y)$ is, clearly, exactly the one in (\ref{56}), with the same
condition on the ratio $\frac{\omega_+}{\omega_-}$ as before. The wave-function $\varphi_{0,0}(x,y)$ belongs to $\Hil_1$ if the following
inequalities are satisfied: \be c_1+\frac{\beta\omega_+}{\Gamma}>0,\qquad c_2-\frac{\delta}{\alpha\omega_+}>0. \label{72}\en Notice also that $\varphi_{0,0}(x,y)$ is eigenvector of $H$ with eigenvalue $\frac{1}{2}\left(\omega_++\omega_-\right)$. This may appear as
an improvement with respect to the result of the previous section, since non trivial choices of $c_1$ and $c_2$ for which (\ref{72}) are
satisfied do exist. For any such choice $\varphi_{0,0}(x,y)$ belongs to $\Hil_1$ and it also belongs to the domain of all the powers of $b_-$
and $b_+$, so that the wave-functions $\varphi_{n_+,n_-}(x,y)$ can be defined as in (\ref{22}). Let us now look for the function
$\Psi_{0,0}(x,y)$. Due to (\ref{71}) we get \be
\Psi_{0,0}(x,y)=N_\Psi\,\exp\left\{-\,\frac{\beta\,\omega_+}{2\,\Gamma}\,x^2+\frac{\delta}{2\,\alpha\,\omega_+}\,
y^2\right\}\exp\left\{c_1x^2+c_2y^2\right\}, \label{73}\en which belongs to $\Hil_1$ if \be \frac{\beta\omega_+}{\Gamma}-c_1>0,\qquad
c_2+\frac{\delta}{\alpha\omega_+}<0. \label{74}\en  $\Psi_{0,0}(x,y)$ is eigenvector of $H^*$, which is different from $H^\dagger=H$, with eigenvalue $\frac{1}{2}\left(\omega_++\omega_-\right)$. It is not hard to check, now, that these conditions are not compatible with those in
(\ref{72}): in other words, it is not possible to fix $c_1$ and $c_2$ in such a way both (\ref{72}) and (\ref{74}) are satisfied. This means
that our original simple-minded idea that adding a weight in the scalar product of the Hilbert space should {\em regularize} the situation does not
work as expected. More explicitly, if $c_1$ and $c_2$ satisfy (\ref{72}), then $\Psi_{0,0}(x,y)$ does not satisfy Assumption 2. Viceversa, if they satisfy
(\ref{74}), then $\varphi_{0,0}(x,y)$ does not satisfy Assumption 1. In both cases, therefore, only a single set of functions in $\Hil_1$ can
be constructed, which is (most likely) a basis of $\Hil_1$ itself.

\vspace{2mm} We refer to \cite{DHO} for more results on this system, which again suggest that pseudo-bosonic wave-functions cannot belong to the Hilbert space in which the model lives independently of the choice of the scalar product, its related adjoint, or of the representation of the pseudo-bosonic operators. This apparently negative result has been considered in \cite{DHO} deeply connected to the dissipative nature of the quantum system under analysis.

\section{Conclusions}

In this paper we have reviewed some results of ours on pseudo-bosons, both from a mathematical and on a physical side. We have seen that few reasonable assumptions produce a rather rich mathematical structure: biorthogonal sets and, sometime, even Riesz bases. We have also seen that some models relevant in pseudo-hermitian quantum mechanics fit well in our framework.

Future works will consider possible new physical models and further mathematical results. We also plan to extend some of the structure of Section II to hamiltonians which are not necessarily of the form $H=BA$, with $A$ and $B$ pseudo-bosonic operators, and whose eigenvalues $E_n$ are not necessarily linear in $n$.

\section*{Acknowledgements}
   The author acknowledge M.I.U.R. for financial support.

\end{document}